\documentclass[10pt,journal,compsoc]{IEEEtran}
\usepackage[nocompress]{cite}

\newif\iftse
\tsefalse

\newif\ifarchival
\archivaltrue

\usepackage{xparse}
\DeclareDocumentCommand{\rosettagraph}{ m O{3.0} O{2.0}}{\begin{tikzpicture}[align=center, xscale=#2,yscale=#3]\input{#1.tikz.tex}\end{tikzpicture}}
\usepackage{dcolumn}
\usepackage{subcaption}
\newcommand{\myparagraph}[1]{\textbf{#1}}

\title{Bayesian Data Analysis in \\Empirical Software Engineering Research}
\author{Carlo A.~Furia, Robert Feldt, and Richard Torkar%
  \IEEEcompsocitemizethanks{%
    \IEEEcompsocthanksitem C.A. Furia is with the Software Institute, Faculty of Informatics, Universit\`a della Svizzera italiana (USI), Lugano, Switzerland. E-mail: see \url{bugcounting.net}
    \IEEEcompsocthanksitem R. Feldt and R. Torkar are with the Software Engineering Division, Department of Computer Science and Engineering, Chalmers University of Technology and the University of Gothenburg, Gothenburg, Sweden.}
  \iftse
\thanks{Manuscript received 14 December 2018; revised 14 July 2019; accepted 13 August 2019.}\fi}
      
\makeatletter
\let\Title\@title
\makeatother

\usepackage[pdftex]{hyperref}
\hypersetup{
pdftitle={\Title{}},
pdfauthor={Furia, Feldt, Torkar},
pdfsubject={},
pdfkeywords={},
pdfcreator={pdfTeX}
}

\usepackage{url}

\usepackage{mathpazo}
\usepackage[T1]{fontenc} 
\usepackage[utf8]{inputenc}

\usepackage{amsmath,amssymb,amsthm}
\DeclareMathOperator{\sgn}{sgn}
\newcommand*\de{\mathop{}\!\mathrm{d}}
\theoremstyle{definition}
\newtheorem{example}{Example}[section]

\usepackage{amsfonts}
\newcommand{\prob}{\mathbb{P}}
\newcommand{\like}{\mathcal{L}}
\newcommand{\prior}{\pi}
\newcommand{\post}{\mathcal{P}}
\newcommand{\ci}[2][95]{\ensuremath{\mathcal{C}_{{#1}\%}^{\shortvar{#2}}}}
\newcommand{\cilow}{\ell}
\newcommand{\cihigh}{h}
\newcommand{\pmf}{p.m.f.\xspace}
\newcommand{\pdf}{p.d.f.\xspace}
\newcommand{\correct}{\textsc{ok}}
\newcommand{\wrong}{\textsc{err}}

\usepackage{proof}

\usepackage{booktabs}
\usepackage{xcolor}
\definecolor{sweak}{RGB}{5,113,176}
\definecolor{smedium}{RGB}{5,113,176}
\definecolor{sstrong}{RGB}{202,0,32} 

\definecolor{lang1}{RGB}{117,112,179}
\definecolor{lang2}{RGB}{27,158,119}

\usepackage{rotating}
\usepackage{multirow}

\usepackage{enumerate}
 \usepackage[inline]{enumitem}

\usepackage{changepage} 

\ifarchival
\usepackage{tikzexternal}
\else
\usepackage{tikz}
\tikzstyle{lang}=[draw=none,font=\footnotesize,inner sep=1pt,outer sep=1pt]
\usepackage{pgfplots}
\pgfmathdeclarefunction{gauss}{2}{  \pgfmathparse{1/(#2*sqrt(2*pi))*exp(-((x-#1)^2)/(2*#2^2))}}

\usetikzlibrary{external}
\fi

\usepackage{array}
\newcommand{\up}{\ensuremath{\mathcal{U}}}
\newcommand{\np}{\ensuremath{\mathcal{N}}}
\newcommand{\spd}{\ensuremath{\mathcal{S}}}
\newcommand{\normal}[2]{\ensuremath{\mathcal{N}(#1, #2)}}
\newcommand{\dist}[1]{\ensuremath{\mathcal{#1}}}

\usepackage{xspace}
\newcommand{\autotest}{Ceccato et al.~\cite{autotest-bib}\xspace}
\newcommand{\rosetta}{Nanz and Furia~\cite{rosetta}\xspace}
\newcommand{\bench}{Bench\xspace}

\usepackage{xstring}
\DeclareRobustCommand{\var}[1]{\ensuremath{\textnormal{\textsl{#1}}}}
\DeclareRobustCommand{\shortvar}[1]{\ensuremath{\textnormal{\textsc{\StrLeft{#1}{1}}}}}
\DeclareRobustCommand{\coef}[1]{\ensuremath{c_{\shortvar{#1}}}}
\DeclareRobustCommand{\estimate}[1]{\ensuremath{\overline{\coef{#1}}}}
\DeclareRobustCommand{\error}[1]{\ensuremath{e_{\shortvar{#1}}}}

\begin{document}

\maketitle

\begin{abstract}
Statistics comes in two main flavors: frequentist and Bayesian.
For historical and technical reasons, frequentist statistics have traditionally dominated empirical data analysis, and
certainly remain prevalent in empirical software engineering.
This situation is unfortunate because frequentist statistics suffer from a number of shortcomings---such as
lack of flexibility and results that are unintuitive and hard to interpret---that curtail their effectiveness
when dealing with the heterogeneous
data that is increasingly available for empirical analysis of software engineering practice.

In this paper, we pinpoint these shortcomings, and present Bayesian data analysis techniques that provide tangible benefits---as they
can provide clearer results that are simultaneously robust and nuanced.
After a short, high-level introduction to the basic tools of Bayesian statistics, we present the reanalysis
of two empirical studies on the effectiveness of automatically generated tests and the performance of programming languages.
By contrasting the original frequentist analyses with our new Bayesian analyses,
we demonstrate the concrete advantages of the latter.
To conclude we advocate a more prominent role for Bayesian statistical techniques in empirical software engineering research and practice.
\end{abstract}

\begin{IEEEkeywords}
Bayesian data analysis, statistical analysis, statistical hypothesis testing, empirical software engineering.
\end{IEEEkeywords}

\section{Introduction}
\label{sec:introduction}

Compared to other fields such as the social sciences,
\emph{empirical software engineering} has been using standard statistical practices for a relatively short period of time~\cite{Bayes1}.
In light of the replication crisis mounting in several of the more established research fields~\cite{replication-crisis},
empirical software engineering's relative statistical immaturity can actually be an opportunity 
to embrace more powerful and flexible statistical methods---which can elevate the impact of the whole field and the solidity of its research findings.

In this paper, we focus on the difference between so-called \emph{frequentist} and \emph{Bayesian} statistics.
For historical and technical reasons~\cite{belief-icse16,fienberg2006,rao1992}, frequentist statistics are the most widely used, and the customary choice in empirical software engineering (see \autoref{sec:relatedWork}) as in other experimental research areas.
In recent years, many statisticians have repeatedly pointed out~\cite{pvalue-cohen,all-false}
that frequentist statistics suffer from a number of shortcomings, 
which limit their scope of applicability and usefulness in practice, and may even lead to drawing flat-out unsound conclusions in certain contexts.
In particular, the widely popular techniques for null hypothesis statistical testing---based on computing the infamous $p$-values---have been \emph{de facto} deprecated~\cite{misusingPval,insignificance-pval},
but are still routinely used by researchers who simply lack practical alternatives: 
techniques that are rigorous yet do not require a wide statistical know-how, and are fully supported by easy-to-use flexible analysis tools.

Bayesian statistics has the potential to replace frequentist statistics, addressing most of the latter's intrinsic shortcomings, 
and supporting detailed and rigorous statistical analysis of a wide variety of empirical data.
To illustrate the effectiveness of Bayesian statistics in practice, 
we present two \emph{reanalyses} of previous work in empirical software engineering.
In each reanalysis, we first describe the original data;
then, we replicate the findings using the same frequentist techniques used in the original paper;
after pointing out the blind spots of the frequentist analysis,
we finally perform an internal replication~\cite{Natalia-replication} that analyzes the same data using Bayesian techniques---leading
to outcomes that are clearer and more robust.
The bottom line of our work is not that frequentist statistics are intrinsically unreliable---in fact, if properly applied in some conditions,
they might lead to results that are close to those brought by Bayesian statistics~\cite{multiple-perspectives}---but that they are generally less intuitive, depend on subtle assumptions, and are thus easier to misuse (despite looking simple superficially).

Our first reanalysis, presented in \autoref{sec:testing}, targets data about debugging using manually-written vs.\ automatically-generated tests.
The frequentist analysis by \autotest is based on (frequentist) linear regression, 
and is generally sound and insightful.
However, since it relies on frequentist statistics to assess statistical significance,
it is hard to consistently map some of its findings to practical significance 
and, more generally,
to interpret the statistics in the quantitative terms of the application domain.
By simply switching to Bayesian linear regression,
we show many of these issues can be satisfactorily addressed:
Bayesian statistics estimate actual probability distributions of the measured data,
based on which one can readily assess practical significance 
and even build a \emph{prediction model} to be used for estimating 
the outcome of similar experiments, 
which can be fine-tuned with any new experimental data.

The second reanalysis, presented in \autoref{sec:rosetta}, targets data about the performance of different programming languages implementing algorithms in the Rosetta Code repository~\cite{rosettacode}.
The frequentist analysis by \rosetta is based on a combination of null hypothesis testing and effect sizes,
which establishes several interesting results---but 
leads to inconclusive or counterintuitive outcomes in some of the language comparisons:
applying frequentist statistics sometimes seems 
to reflect idiosyncrasies of the experimental data rather than intrinsic differences between languages.
By performing full-fledged Bayesian modeling,
we infer a more coherent picture of the performance differences between programming languages:
besides the advantage of having actual estimates of average speedup of one language or another,
we can quantitatively analyze the \emph{robustness} of the analysis results,
and point out what can be improved by collecting additional data in new experiments.

\myparagraph{Contributions.}
This paper makes the following contributions:
\begin{itemize}
\item a discussion of the shortcomings of the most common frequentist statistic techniques often used in empirical software engineering;
\item a high-level introduction to Bayesian analysis;
\item detailed analyses of two case studies \cite{autotest-bib,rosetta}
  from empirical software engineering,
  discussing how the limitations of frequentist statistics weaken clarity and generality
  of the results, while Bayesian statistics make for robust and rich analyses.
\end{itemize}

\myparagraph{Prerequisites.}
This paper only requires a basic familiarity with the fundamental notions of probability theory.
Then, \autoref{sec:bayes-overview} introduces the basic principles of Bayesian statistics
(including a concise presentation of Bayes' theorem), and illustrates how frequentist techniques
such as statistical hypothesis testing work and the shortcoming they possess.
Our aim is that our presentation be accessible to a majority of empirical software engineering researchers.

\myparagraph{Scope.}
This paper's main goal is demonstrating
the shortcomings of commonly used frequentist techniques,
and to show how Bayesian statistics could be a better choice for data analysis.
This paper is \emph{not} meant to be:
\begin{enumerate}
\item A tutorial on applying Bayesian statistic tools;   we refer readers to textbooks~\cite{BDA,puppies,rethinking} for such step-by-step introductions.

\item A philosophical comparison of frequentist vs.\ Bayesian interpretation of statistics.

\item A criticism of the two papers (\autotest and \rosetta)
  whose frequentist analyses we peruse in \autoref{sec:testing} and \autoref{sec:rosetta}.
  We chose those papers because they carefully apply generally accepted best practices,
  in order to demonstrate that, even when applied properly, frequentist statistics have
  limitations and bring results that can be practically hard to interpret.
\end{enumerate}

\myparagraph{Availability.}
All machine-readable data and analysis scripts used in this paper's analyses are available online at
\begin{center}
  \url{https://bitbucket.org/caf/bda-in-ese}
\end{center}

\subsection{Related Work}
\label{sec:relatedWork}

\myparagraph{Empirical research in software engineering.}
Statistical analysis of empirical data has become commonplace in software engineering research~\cite{experiments-book,hitchhiker-icse11,Bayes1}, and it is even making its way into software development practices~\cite{data-scientist}.

As we discuss below, the overwhelming majority of statistical techniques that are being used in software engineering empirical research are, however, of the frequentist kind, with Bayesian statistics hardly even mentioned.

Of course, Bayesian statistics is a fundamental component of many machine learning techniques~\cite{ml-book,ml-springer}; as such, it is used in software engineering research indirectly whenever machine learning is used.
In this paper, however, we are concerned with the direct usage of statistics to analyze empirical data from the scientific perspective---a pursuit that seems mainly confined to frequentist techniques in software engineering~\cite{Bayes1}.
As we argue in the rest of the paper, this is a lost opportunity because Bayesian techniques do not suffer from several limitations of frequentist ones, and can support rich, robust analyses in several situations.

\myparagraph{Bayesian analysis in software engineering?}
To validate the impression that Bayesian statistics are not normally used in empirical software engineering, we carried out a small literature review of ICSE papers.\footnote{\cite{views-icse15} has a much more extensive literature survey of empirical publications in software engineering.}
We selected all papers from the main research track of the latest six editions of the International Conference on Software Engineering (ICSE 2013 to ICSE 2018) that mention ``empirical'' in their title or in their section's name in the proceedings.
This gave 25 papers, from which we discarded one~\cite{stochastic-icse14} that turned out not to be an empirical study.
The experimental data in the remaining 24 papers come from various sources: 
the output of analyzers and other tools~\cite{mutations-icse17,compilers-icse16,browser-icse13,configuration-icse14,equivalence-icse15}, 
the mining of repositories of software and other artifacts~\cite{evolving-icse14,javareflection-icse17,verification-icse15,js-icse16,fixing-icse13,fixes-icse15}, 
the outcome of controlled experiments involving human subjects~\cite{summaries-icse16,lambdas-icse16,smells-icse13},
interviews and surveys~\cite{codereviews-icse13,coupling-icse13,belief-icse16,network-icse17,data-icse16,green-icse16,uml-icse13,views-icse15},
and a literature review~\cite{grounded-icse16}.

As one would expect from a top-tier venue like ICSE, the 24 papers follow recommended practices in reporting and analyzing data, using significance testing (6 papers), effect sizes (5 papers), correlation coefficients (5 papers), frequentist regression (2 papers), and visualization in charts or tables (23 papers).
None of the papers, however, uses Bayesian statistics.
In fact, no paper but two~\cite{belief-icse16,fixing-icse13} even mentions the terms ``Bayes'' or ``Bayesian''.
One of the exceptions~\cite{fixing-icse13} only cites Bayesian machine-learning techniques used in related work to which it compares.
The other exception~\cite{belief-icse16} includes a presentation of the two views of frequentist and Bayesian statistics---with a critique of $p$-values similar to the one we make in \autoref{sec:classic-vs-bayes}---but does not show how the latter can be used in practice.
The aim of \cite{belief-icse16} is investigating the relationship between empirical findings in software engineering and the actual beliefs of programmers about the same topics.
To this end, it is based on a survey of programmers whose responses are analyzed using frequentist statistics; Bayesian statistics is mentioned to frame the discussion about the relationship between evidence and beliefs, but does not feature past the introductory second section.
Our paper has a more direct aim: to concretely show how Bayesian analysis can be applied in practice in empirical software engineering research, as an alternative to frequentist statistics; thus, its scope is complementary to \cite{belief-icse16}'s.

As additional validation based on a more specialized venue for empirical software engineering,
we also inspected all 105 papers published in Springer's Empirical Software Engineering (EMSE) journal during the year 2018.
Only 22 papers mention the word ``Bayes'':
17 of them refer to Bayesian machine learning classifiers (such as naive Bayes or Bayesian networks);
2 of them discuss Bayesian optimization algorithms for machine learning (such as latent Dirichlet allocation word models);
3 of them mention ``Bayes'' only in the title of some bibliography entries.
None of them use Bayesian statistics as a replacement of classic frequentist data analysis techniques.

More generally, we are not aware of any direct application of Bayesian data analysis to empirical software engineering data with the exception of \cite{bayes-extended,F-ICSE17-poster} and \cite{Ernst-bayes}.
The technical report~\cite{bayes-extended} and its short summary~\cite{F-ICSE17-poster}
are our preliminary investigations along the lines of the present paper.
Ernst~\cite{Ernst-bayes} presents a conceptual replication of an existing study to argue the analytical effectiveness of multilevel Bayesian models.

\myparagraph{Criticism of the $p$-value.}
Statistical hypothesis testing---and its summary outcome, the $p$-value---has been customary in experimental science for many de\-cades, both for the influence~\cite{rao1992} of its proponents Fisher, Neyman, and Pearson, and because it offers straightforward, ready-made procedures that are computationally simple\footnote{With modern computational techniques it is much less important whether statistical methods are computationally simple; we have CPU power to spare---especially if we can trade it for stronger scientific results.}.
More recently, criticism of frequentist hypothesis testing has been voiced in many experimental sciences, such as psychology~\cite{pvalue-cohen,pvalue-psychology} and medicine~\cite{pvalue-medicine}, that used to rely on it heavily, as well as in statistics research itself~\cite{ASA-statement,pvalue-statisticians,gelman-pvalues}.
The criticism, which we articulate in \autoref{sec:classic-vs-bayes}, concludes that $p$-value-based hypothesis testing should be abandoned~\cite{abandon-significance,riseup}.\footnote{Even statisticians who still accept null-hypothesis testing recognize the need to change the way it is normally used~\cite{down-to-005}.} There has been no similar explicit criticism of $p$-values in software engineering research, and in fact statistical hypothesis testing is still regularly used~\cite{Bayes1}.

\myparagraph{Guidelines for using statistics.}
Best practices of using statistics in empirical software engineering are described in books~\cite{datascience-perspectives,experiments-book}
and articles~\cite{hitchhiker-icse11,JedlitschkaJR14}.
Given their focus on frequentist statistics,\footnote{\cite{hitchhiker-icse11,JedlitschkaJR14,experiments-book} do not mention Bayesian techniques; \cite{hitchhiker-journal} mentions their existence only to declare they are out of scope; one chapter~\cite{bayes-in-book} of~\cite{datascience-perspectives} outlines Bayesian networks as a machine learning technique.} they all are complementary to the present paper, whose main goal is showing how Bayesian techniques can add to, or replace, frequentist ones, and how they can be applied in practice.

\section{An Overview of Bayesian Statistics}
\label{sec:bayes-overview}

Statistics provides models of \emph{events}, such as the output of a randomized algorithm; the probability function $\prob$ assigns probabilities---values in the real unit interval $[0, 1]$, or equivalently percentages in $[0, 100]$---to events.
Often, events are the values taken by \emph{random variables} that follow a certain probability \emph{distribution}.
For example, if $X$ is a random variable modeling the throwing of a six-face dice, it means that $\prob[x] = 1/6$ for $x \in [1..6]$, and $\prob[x] = 0$ for $x \not\in [1..6]$---where $\prob[x]$ is a shorthand for $\prob[X = x]$, and $[m..n]$ is the set of integers between $m$ and $n$.

The probability of variables over discrete domains is described by \emph{probability mass functions} (\pmf for short); their counterparts over continuous domains are probability density functions (\pdf), whose integrals give probabilities.
The following presentation mostly refers to continuous domains and \pdf,
although notions apply to discrete-domain variables as well with a few technical differences.
For convenience, we may denote a distribution and its \pdf with the same symbol; for example, random variable $X$ has a \pdf also denoted $X$, such that $X[x] = \prob[x] = \prob[X = x]$.

\myparagraph{Conditional probability.}
The \emph{conditional} probability $\prob[h \mid d]$ is the probability of $h$ given that $d$ has occurred.
For example, $d$ may represent the empirical \emph{data} that has been \emph{observed}, and $h$ is a \emph{hypothesis} that is being tested.

Consider a static analyzer that outputs $\top$ (resp.\ $\bot$) to indicate that the input program never overflows (resp.\ may overflow); $\prob[\correct \mid \top]$ is the probability that, when the algorithm outputs $\top$, the input program is indeed free from overflows---the data is the output ``$\top$'' and the hypothesis is ``the input does not overflow''.

\subsection{Bayes' Theorem}
\label{sec:bayes-theorem}

\myparagraph{Bayes' theorem}
connects the conditional probabilities \linebreak$\prob[h \mid d]$ (the probability that the hypothesis is correct given the experimental data)
and $\prob[d \mid h]$ (the probability that we would see this experimental data given that the hypothesis actually was true).
The famous theorem states that
\begin{equation}
\prob[h \mid d]
\ =\ 
\frac{\prob[d \mid h]\cdot\prob[h]}{\prob[d]}\,.
\label{eq:bayes-th}
\end{equation}
Here is an example of applying Bayes' theorem.
\begin{example}\label{ex:static-analyzer}
Suppose that the static analyzer gives true positives and true negatives with high probability ($\prob[\top \mid \correct] = \prob[\bot \mid \wrong] = 0.99$), and that many programs are affected by some overflow errors ($\prob[\correct] = 0.01$).
Whenever the analyzer outputs $\top$, what is the chance that the input is indeed free from overflows?
Using Bayes' theorem, $\prob[\correct \mid \top] = (\prob[\top \mid \correct]\cdot \prob[\correct])/\prob[\top] = (\prob[\top \mid \correct]\cdot \prob[\correct])/(\prob[\top \mid \correct]\cdot \prob[\correct] + \prob[\top \mid \wrong]\, \prob[\wrong]) = (0.99 \cdot 0.01)/(0.99 \cdot 0.01 + 0.01 \cdot 0.99) = 0.5$, we conclude that we can have a mere 50\% confidence in the analyzer's output.
\end{example}

\myparagraph{Priors, likelihoods, and posteriors.}
In Bayesian analysis~\cite{thinkBayes}, each factor of \eqref{eq:bayes-th} has a special name:
\begin{enumerate}
\item $\prob[h]$ is the \emph{prior}---the probability of the hypothesis $h$ before having considered the data---written $\prior[h]$;
\item $\prob[d \mid h]$ is the \emph{likelihood} of the data $d$ under hypothesis $h$---written $\like[d; h]$;
\item $\prob[d]$ is the \emph{normalizing constant};
\item and $\prob[h \mid d]$ is the \emph{posterior}---the probability of the hypothesis $h$ after taking data $d$ into account---written $\post_d[h]$.
\end{enumerate}
With this terminology, we can state Bayes' theorem \eqref{eq:bayes-th} as ``\emph{the posterior is proportional to the likelihood times the prior}'', that is,

\begin{equation}
  \prob[h \mid d] \propto \prob [d \mid h] \times \prob [h]\,,
  \label{eq:propto}
\end{equation}
and hence we \emph{update} the prior to get the posterior.

The only role of the normalizing constant is ensuring that the posterior defines a correct probability distribution when evaluated over all hypotheses.
In most cases we deal with hypotheses $h \in H$ that are mutually exclusive and exhaustive; then, the normalizing constant is simply $\prob[d] = \int_{h \in H} \prob[d \mid h]\,\prob[h] \de h$, which can be computed from the rest of the information.
Thus, it normally suffices to define likelihoods that are \emph{proportional to} a probability, and rely on the \emph{update rule} to normalize them and get a proper probability distribution as posterior.

\subsection{Frequentist vs.\ Bayesian Statistics}
\label{sec:classic-vs-bayes}

Despite being a simple result about an elementary fact in probability, Bayes' theorem has significant implications for statistical reasoning. We do not discuss the philosophical differences between how frequentist and Bayesian statistics interpret their results~\cite{philosophy}.
Instead, we focus on describing how some features of Bayesian statistics support new ways of analyzing data.
We start by criticizing statistical hypothesis testing since it is a customary technique in frequentist statistics that is widely applied in empirical software engineering research, and suggest how Bayesian techniques could provide more reliable analyses.
For a systematic presentation of Bayesian statistics see Kruschke~\cite{puppies}, McElreath~\cite{rethinking}, and Gelman et al.~\cite{BDA}.

\subsubsection{Hypothesis Testing vs.\ Model Comparison}
\label{sec:NHST}
A primary goal of experimental science is validating models of behavior based on empirical data.
This often takes the form of choosing between alternative hypotheses, such as, in the software engineering context, deciding whether automatically generated tests support debugging better than manually written tests (\autoref{sec:testing}), or whether a programming language is faster than another (\autoref{sec:rosetta}).
 
\emph{Hypothesis testing} is the customary framework offered by frequentist statistics to choose between hypotheses.
In the classical setting, a null hypothesis $H_0$ corresponds to ``no significant difference'' between two \emph{treatments} $A$ and $B$ (such as two static analysis algorithms whose effectiveness we want to compare); an alternative hypothesis $H_1$ is the null hypothesis's negation, which corresponds to a significant difference between applying $A$ and applying $B$.
A \emph{null hypothesis significance test}~\cite{hitchhiker-journal}, such as the $t$-test or the $U$-test, is a procedure that takes as input a combination $D^*$ of two datasets $D_A$ and $D_B$, respectively recording the outcome of applying $A$ and $B$, and outputs a probability called the $p$-value.

\begin{figure}[t]
  \centering
  \includegraphics[width=0.44\textwidth]{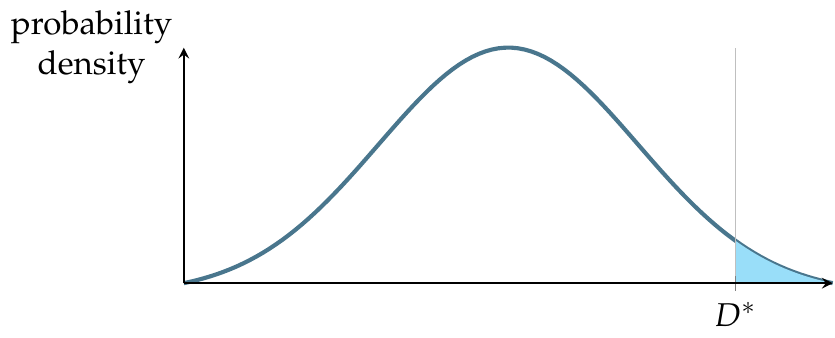}
\caption{The $p$-value is the probability, under the null hypothesis, of drawing data that is at least as extreme as the observed data $D^*$.
  Graphically, the $p$-value is the shaded area under the curve, which models the probability distribution under the null hypothesis.}
\label{fig:pvalue-def}
\end{figure}

The $p$-value is the probability, under the null hypothesis, of
observing data at least as extreme as the ones that were actually observed. Namely, it is the probability $\prob[D \geq D^* \mid H_0]$\footnote{Depending on the nature of the data, the $p$-value may also be defined as a left-tail event $\prob[D \leq D^* \mid H_0]$ or as a 2-tailed event $2 \cdot \min \left(\prob[D \leq D^* \mid H_0], \prob[D \geq D^* \mid H_0]\right)$.}
of drawing data $D$ that is equal to the observed data $D^*$ or more ``extreme'',
conditional on the null hypothesis $H_0$ holding.
As shown visually in \autoref{fig:pvalue-def},
data is expected to follow a certain distribution under the null hypothesis (the actual distribution depends on the statistical test that is used);
the $p$-value measures the tail probability of drawing data equal to the observed $D^*$ or more unlikely than it---in other words, 
how far the observed data is from the most likely observations under the null hypothesis.
If the $p$-value is sufficiently small---typically $p \leq 0.05$ or $p \leq 0.01$---it means that the null hypothesis is an unlikely explanation of the observed data.
Thus, one \emph{rejects} the null hypothesis, which corresponds to leaning towards preferring the alternative hypothesis $H_1$ over $H_0$:
in this case, we have increased our confidence that $A$ and $B$ differ.

Unfortunately, this widely used approach to testing hypotheses suffers from serious shortcomings~\cite{pvalue-cohen}, which have prompted calls to seriously reconsider how it's used as a statistical practice~\cite{ASA-statement,down-to-005}, or even to abandon it altogether~\cite{abandon-significance,riseup}.
The most glaring problem is that, in order to decide whether $H_0$ is a plausible explanation of the data, we would need the conditional probability $\prob[H_0 \mid D \geq D^*]$ of the hypothesis given the data, not the $p$-value $\prob[D \geq D^* \mid H_0]$.
The two conditional probabilities are related by Bayes' theorem \eqref{eq:bayes-th}, but knowing only $\prob[D \geq D^* \mid H_0]$ is not enough to determine $\prob[H_0 \mid D \geq D^*]$;\footnote{Assuming that they are equal is the ``confusion of the inverse''~\cite{fallacy-inverse}.} in fact, \autoref{sec:bayes-theorem}'s example of the static analyzer (\autoref{ex:static-analyzer}) showed a case where one conditional probability is 99\% while the other is only 50\%.

Rejecting the null hypothesis in response to a small $p$-value looks like a sound inference,
but in reality it is not, because it follows an unsound probabilistic extension
of a reasoning rule that is perfectly sound in Boolean logic.
The sound rule is \emph{modus tollens}:
\begin{align}
  \begin{array}{c}
    \text{if }X\text{ implies that }Y\text{ is false} \\
    \text{and we observe }Y\\
    \text{then }X\text{ is false}
  \end{array}
  &&
  \infer{\neg X}{X \longrightarrow \neg Y & Y}
  \label{eq:modus-tollens}
\end{align}
For example, if $X$ means ``a person lives in Switzerland'' and $Y$ means ``a person is the King of Sweden'', if we observe a person that is the Swedish king we can conclude that he does not live in Switzerland.
The probabilistic extension of modus tollens, which is unsound, is:
\begin{align}
  \begin{array}{c}
    \text{if }X\text{ implies that }Y\text{ is \emph{probably} false} \\
    \text{(equivalently: }Y\text{ is improbably true)} \\
    \text{and we observe }Y\\
    \text{then }X\text{ is \emph{probably} false}
  \end{array}
  &&
  \infer{\prob[X] < \epsilon}{\prob[Y \mid X] < \epsilon & Y}
  \label{eq:modus-tollens-prob}
\end{align}
To see that rule \eqref{eq:modus-tollens-prob} is unsound let's consider the case where $Y$ means ``a person is the Queen of England'' and $X$ means ``a person lives in London''.
Clearly, $\prob[Y \mid X]$ is small because only one person out of million Londoners is indeed the Queen of England.
However, if the observed $Y$ happens to be the Queen of England, we would be wrong to infer that she's unlikely to live in London; indeed, the sound conclusion is that she \emph{definitely} lives in London.
Whenever we reject $X = H_0$ after observing $Y = D$ following a calculation that $\prob[Y \mid X] = p$ is small, we are applying this unsound rule.

This unsound inference is not the only issue with using $p$-values.
Methodological problems come from how hypothesis testing pits the null hypothesis against the alternative hypothesis: as the number of observations grows, it becomes increasingly likely that \emph{some} effect is detectable (or, equivalently, it becomes increasingly unlikely that \emph{no effects} are), which leads to rejecting the null hypothesis, independent of the alternative hypothesis, just because it is unreasonably restrictive (that is, it's generally unlikely that there are genuinely no effects).
This problem may result both in suggesting that some negligible effect is significant just because we reject the null hypothesis,
and in discarding some interesting experimental results just because they fail to trigger the arbitrary $p < 0.05$ threshold of significance.
This is part of the more general problem with insisting on a dichotomous view between two alternatives: a better approach would be based on richer statistics than one or few summary values (such as the $p$-value) and would combine quantitative and qualitative data (for example, about the sign and magnitude of an effect) to get richer pictures.

To sum up, null hypothesis testing has both technical problems ($p$-values are insufficient statistics to conclude anything about the null hypothesis)
and methodological ones (null hypotheses often are practically useless models).

\subsubsection{Types of Statistical Inference Errors}
\label{sec:type12-sm-errors}
Frequentist statistics often analyze the probability of so-called \emph{type 1} and \emph{type 2} errors:
\begin{description}
\item[Type 1:] a type 1 error is a \emph{false positive}: rejecting a hypothesis when it is actually true;
\item[Type 2:] a type 2 error is a \emph{false negative}: accepting a hypothesis when it is actually false.
\end{description}

In the context of hypothesis testing, the probability of a type 1 error is related to the $p$-value,
but the $p$-value alone does \emph{not} ``denote the probability of a Type~I error''~\cite{hitchhiker-journal}:
the $p$-value is a probability conditional on $H_0$ being true,
and hence it cannot be used to conclude whether $H_0$ itself is true---an assessment that is required to estimate the probability of a type~1 error,
which occurs, by definition, only if $H_0$ is true.
In the context of experimental design,
the $p$-value's \emph{rejection threshold} $\alpha$ is an \emph{upper bound} on the probability of a type~1 error~\cite{pval-misconceptions}.
In this case, a type~1 error occurs when $H_0$ is true but $p < \alpha$; in order to assign a probability to this scenario, we need to assign a \emph{prior probability} that the null hypothesis is true: \begin{equation}
  \label{eq:type1prob}
  \prob[\text{type~1 error}] = \prob[ p < \alpha \mid H_0] \cdot \prob[H_0] = \alpha \cdot \prob[H_0]
\end{equation}
Therefore, the probability of a type 1 error in an experiment with rejection threshold set to $\alpha$ lies between zero and $\alpha$.\footnote{You will still find the erroneous statement that $\alpha$ is the probability of a type~1 error in numerous publications including some bona fide statistics textbooks~\cite{warner2012applied,reid2013introduction}.}

Note, however, that even the technically correct interpretation of the significance level $\alpha$ as an upper bound on the probability of a type 1 error
breaks down when we have \emph{multiple testing}, as we illustrate in \autoref{sec:rosetta-multiple}.
In all cases, the $p$-value is a statistics of the data alone, and hence it is insufficient to estimate the probability of a hypothesis being true.

Bayesian analysis prefers to avoid the whole business of hypothesis testing
and instead computing actual posterior probability distributions based on data and priors.
The Bayesian viewpoint also suggests that the usual frequentist preoccupation with type 1 and 2 errors is often misplaced:
the null hypothesis captures a narrow notion of ``no effects'', which is rarely the case.
As Gelman remarks:\footnote{Andrew Gelman's blog \href{https://andrewgelman.com/2004/12/29/type_1_type_2_t/}{\emph{Type 1, type 2, type S, and type M errors}}, 29 December 2004.}
\begin{quotation}
  A type~1 error occurs only if the null hypothesis is true (typically if a certain parameter, or difference in parameters, equals zero).
  In the applications I've worked on [\ldots]
  I've never come across a null hypothesis that could actually be true. A type~2 error occurs only if I claim that the null hypothesis is true, and I would certainly not do that, given my statement above!
\end{quotation}

Instead, Gelman recommends~\cite{TypeSM} focusing on type $S$ and type $M$ errors. A type $S$ error occurs when we mistakenly infer the \emph{sign} of an effect:
for example that a language $A$ is consistently faster than language $B$, when in fact it is B that is faster than A.
A type $M$ error occurs when we overestimate the \emph{magnitude} of an effect:
that language $A$ is 3 times faster than language $B$, when in fact it is at most 5\% faster.
\autoref{sec:rosetta:bayes-results} discusses type S and type M errors in \rosetta's case study.

\subsubsection{Scalar Summaries vs.\ Posterior Distributions}
\label{sec:scalar-vs-posterior}
Even though hypothesis testing normally focuses on two complementary hypotheses ($H_0$ and its negation $H_1$),
it is often necessary to simultaneously compare several alternative hypotheses $h \in H$ grouped in a discrete or even dense set $H$.
For example, in \autoref{sec:rosetta}'s reanalysis, $H$ is the continuous interval of all possible \emph{speedups} of one language relative to another.
A distinctive advantage of full-fledged Bayesian statistics is that it supports deriving a complete \emph{distribution} of posterior probabilities, by applying \eqref{eq:bayes-th} for all hypotheses $h \in H$, rather than just scalar summaries (such as estimators of mean, median, and standard deviation, standardized measures of effect size, or $p$-values on discrete partitions of $H$). Given a distribution we can still compute scalar summaries (including confidence\slash credible intervals), but we retain additional advantages such as being able to visualize the distribution and to derive other distributions by iterative application of Bayes' theorem.
This supports decisions based on a variety of criteria and on a richer understanding of the experimental data, as we demonstrate in the case studies of \autoref{sec:testing} and \autoref{sec:rosetta}.

\subsubsection{The Role of Prior Information}
\label{sec:role-of-priors}
The other distinguishing feature of Bayesian analysis is that it \emph{starts from a prior probability} which models the initial knowledge about the hypotheses.
The prior can record previous results in a way that is congenial to the way science is supposed to work---not as completely independent experiments in an epistemic vacuum, but by constantly scrutinizing previous results and updating our models based on new evidence.

A canned criticism of Bayesian statistics observes that using a prior is a potential source of bias.
However, explicitly taking into account this very fact helps analyses be more rigorous and more comprehensive.
In particular, we often consider several different alternative priors to perform Bayesian analysis.
Priors that do not reflect any strong a priori bias are traditionally called \emph{uninformative};
a uniform distribution over hypotheses is the most common example.
Note, however, that the term ``uninformative'' is misleading:
a uniform prior encodes as much information as a non-uniform one---it is just a different kind of information.\footnote{In fact, in a statistical inference context, a uniform prior often leads to overfitting because we learn too much from the data without correction~\cite{rethinking}.} 
In a different context, this misunderstanding is illustrated metaphorically by this classic \emph{koan}~\cite[Appendix A]{HackersDictionary} involving artificial intelligence pioneers Marvin Minsky and Gerald Sussman:
\begin{quotation}
  \noindent
  In the days when Sussman was a novice, Minsky once came to him as he sat hacking at the PDP-6.
  ``What are you doing?'', asked Minsky.
  ``I am training a randomly wired neural net to play Tic-Tac-Toe'' Sussman replied.
  ``Why is the net wired randomly?'', asked Minsky.
  ``I do not want it to have any preconceptions of how to play'', Sussman said.
  Minsky then shut his eyes.
  ``Why do you close your eyes?'', Sussman asked his teacher.
  ``So that the room will be empty.''
  At that moment, Sussman was enlightened.  
\end{quotation}
A uniform prior is the equivalent a randomly wired neural network---with unbiased assumptions but still with specific assumptions.

Whenever the posterior distribution is largely independent of the chosen prior, we say that the data \emph{swamps} the prior, and hence the experimental evidence is decidedly strong.
If, conversely, choosing a suitable, biased prior is necessary to get sensible results, it means that the evidence is not overwhelming, and hence any additional reliable source of information should be vetted and used to sharpen the analysis results.
In these cases, characteristics of the biased priors may suggest what kinds of assumptions about the data distribution we should investigate further.
For example, suppose that sharp results emerge only if
we use a prior that assigns vanishingly low probabilities to a certain parameter $x$ taking negative values.
In this case, the analysis should try to ascertain whether $x$ can or cannot meaningfully take such values:
if we find theoretical or practical reasons why $x$ should be mostly positive, the biased prior acts as  a safeguard against spurious data;
conversely, if $x$ could take negative values in other experiments, we conclude that the results obtained with the biased prior are only valid in a restricted setting, and new experiments should focus on the operational conditions where $x$ may be negative.

Assessing the impact of priors with markedly different characteristics leads to a process called \emph{(prior) sensitivity analysis},
whose goals are clearly described by McElreath~\cite[Ch.~9]{rethinking}:
\begin{quotation}
  A sensitivity analysis explores how changes in assumptions influence inference.
  If none of the alternative assumptions you consider have much impact on inference, that's worth reporting.
  Likewise, if the alternatives you consider do have an important impact on inference, that's also worth reporting.
  [\ldots]
  In sensitivity analysis, many justifiable analyses are tried, and all of them are described.
\end{quotation}
Thus, the goal of sensitivity analysis is \emph{not} choosing a prior, but rather better understanding the statistical model and its features.
\autoref{sec:rosetta:bayes:sensitivity} presents a prior sensitivity analysis for some of the language comparisons of \rosetta's data.

\myparagraph{}
To sum up, Bayesian analysis stresses the importance of careful \emph{modeling} of assumptions and hypotheses, which is more conducive to accurate analyses than the formulaic application of ready-made statistics.
In the rest of the paper, we illustrate how frequentist and Bayesian data analysis work in practice
on two case studies based on software engineering empirical data.
We will see that the strength of Bayesian statistics translates into analyses that are more robust, sound, and whose results are natural to interpret in practical terms.

\subsection{Techniques for Bayesian Inference}
\label{sec:bayesian-inference-algo}

No matter how complex the model, every Bayesian analysis ultimately boils down to computing a \emph{posterior probability distribution}---according to Bayes' theorem \eqref{eq:bayes-th}---given likelihood, priors, and data.
In most cases the posterior cannot be computed \emph{analytically}, but numerical approximation algorithms exist that work as well for all practical purposes.
Therefore, in the remainder of this article we will always mention the ``posterior distribution'' as if it were computed exactly;
provided we use reliable tools to compute the approximations, the fact that we are actually dealing with numerical approximations does not make much practical difference.

The most popular numerical approximation algorithms for Bayesian inference
work by \emph{sampling} the posterior distribution at different points;
for each sampled point $\bar{h}$, we calculate the posterior probability $\post_d[\bar{h}] = \prob[\bar{h} \mid d]$ by evaluating the right-hand side of~\eqref{eq:bayes-th} at $\bar{h}$ for the given likelihood and data $d$.
Sampling must be neither too sparse (resulting in an inaccurate approximation)
nor too dense (taking too much time).
The most widely used techniques achieve a simultaneously effective and efficient sampling by using Markov Chain Monte Carlo (MCMC)~\cite{MCMC-handbook}
randomized algorithms.
In a nutshell, the sampling process corresponds to visiting a Markov chain~\cite[Ch.~6]{FMMR-TimeBook-12},
whose transition probabilities reflect the posterior probabilities in a way that ensures that they are sampled densely enough
(after the chain reaches its steady state).
Tools such as Stan~\cite{Stan-article} provide scalable generic implementations of these techniques.

Users of Bayesian data analysis tools normally need not understand the details of how Bayesian sampling algorithms work.
As long as they use tried-and-true statistical models and follow recommended guidelines (such as those we outline in \autoref{sec:guidelines}),
validity of analysis results should not depend on the details of how the numerical approximations are computed.

\section{Linear Regression Modeling: Significance and Prediction}
\label{sec:testing}

Being able to automatically generate test cases can greatly help discover bugs quickly and effectively: the unit tests generated by tools such as Randoop (using random search~\cite{PachecoE07randoop})
are likely to be complementary to those a human programmer would write, and hence have a high chance of exercising behavior that may reveal flaws in software.
Conversely, the very features that distinguish automatically-generated tests from programmer-written ones---such as uninformative identifiers and a random-looking nature~\cite{readability-autogen}---may make the former less effective than the latter when programmers use tests to \emph{debug} faulty programs.
\autotest designed a series of experiments to study this issue: \emph{do automatically generated tests provide good support for debugging?}

\subsection{Data: Automated vs.\ Manual Testing}
\label{sec:data-testing}

\autotest compare automatically-generated and manually-written tests---henceforth called `autogen' and `manual' tests---according to different features.
The core experiments are empirical studies involving, as subjects, students and researchers, who debugged Java systems based on the information coming from autogen or manual tests.

Our reanalysis focuses on \autotest's first study, 
where autogen tests were produced running \emph{Randoop} 
and the success of debugging was evaluated according to the experimental subjects' \emph{effectiveness}---namely,
how many of the given bugs they could successfully fix given a certain time budget.

Each experimental data point assigns a value to six variables:
\begin{description}
\item[fixed:] the dependent variable measuring the (nonnegative) number of bugs correctly fixed by the subject;
\item[treatment:] \emph{auto} (the subject used autogen tests) or \emph{manual} (the subject used manual tests);
\item[system:] \emph{J} (the subject debugged Java application JTopas) or \emph{X} (the subject debugged Java library XML-Security);
\item[lab:] \emph{1} or \emph{2} according to whether it was the first or second of two debugging sessions each subject worked in;
\item[experience:] \emph{B} (the subject is a bachelor's student) or \emph{M} (the subject is a master's student);
\item[ability:] \emph{low}, \emph{medium}, or \emph{high}, according to a discretized assessment of ability based on the subject's GPA and performance in a training exercise.
\end{description}
For further details about the experimental protocol see \autotest.

\subsection{Frequentist Analysis}
\label{sec:frequentist-testing}

\subsubsection{Linear Regression Model}
\label{sec:autotest:linear-model}
\autotest's analysis fits a linear regression model that expresses outcome variable \var{fixed} 
as a linear combination of five predictor variables:
\begin{multline}
\footnotesize
\var{fixed}
=
\coef{intercept}
+ \coef{treatment} \cdot \var{treatment}
+ \coef{system} \cdot \var{system}
+ \coef{lab} \cdot \var{lab}
\\\footnotesize + \coef{experience} \cdot \var{experience}
+ \coef{ability} \cdot \var{ability}
+ \epsilon
\label{eq:basic-lr-0}
\end{multline}
where each variable's value is encoded on an integer interval scale, 
and $\epsilon$ models the error as a normally distributed random variable with mean zero and unknown variance.
We can rewrite \eqref{eq:basic-lr-0} as:
\begin{equation}
\footnotesize
\var{fixed}
\sim
\dist{N}\left(
  \begin{array}{rl}
\coef{intercept}
& + \coef{treatment} \cdot \var{treatment}
+ \coef{system} \cdot \var{system}
+ \coef{lab} \cdot \var{lab}
\\ & + \coef{experience} \cdot \var{experience}
     + \coef{ability} \cdot \var{ability}
  \end{array}
  ,
\sigma \right)
\label{eq:basic-lr}
\end{equation}

where $\dist{N}(\mu, \sigma)$ is a normal distribution with mean $\mu$ and (unknown) variance $\sigma^2$.
Form \eqref{eq:basic-lr} is equivalent to \eqref{eq:basic-lr-0},
but better lends itself to the generalizations we will explore later on in \autoref{sec:bayesian-testing}.

\begin{table*}[!htbp] \centering 
  \caption{Regression coefficients in the linear model with Gaussian error model \eqref{eq:basic-lr}, computed with frequentist analysis.
    For each coefficient \coef{v}, the table reports
    maximum likelihood \emph{estimate} $\estimate{v}$,
    standard \emph{error} \error{v} of the estimate,
    \emph{$t$ statistic} of the estimate,
    $p$\emph{-value} (the probability of observing the $t$ statistic under the null hypothesis),
    and \emph{lower} and \emph{upper} endpoints of the 95\% confidence interval $\ci{v}$ of the coefficient.
  } 
  \label{tab:mvr-effe-freq-gauss}
  \normalsize
\begin{tabular}{@{\extracolsep{5pt}} D{.}{.}{-3} D{.}{.}{-3} D{.}{.}{-3} D{.}{.}{-3} D{.}{.}{-3} D{.}{.}{-3} D{.}{.}{-3} } 
\toprule \\[-2ex] 
\multicolumn{1}{c}{\textsc{coefficient} \coef{v}} & \multicolumn{1}{c}{\textsc{estimate} $\estimate{v}$} & \multicolumn{1}{c}{\textsc{error} \error{v}} & \multicolumn{1}{c}{$t$ \textsc{statistic}} & \multicolumn{1}{l}{$p$\textsc{-value}} & \multicolumn{1}{c}{\textsc{lower}} & \multicolumn{1}{c}{\textsc{upper}} \\ 
\midrule \\[-1.8ex] 
\multicolumn{1}{l}{\coef{intercept} (intercept)} & -1.756 & 0.725 & -2.422 & 0.020 & -3.206 & -0.306 \\ 
\multicolumn{1}{l}{\coef{treatment} (treatment)} & 0.651 & 0.330 & 1.971 & 0.055 & -0.010 & 1.311 \\ 
\multicolumn{1}{l}{\coef{system} (system)} & -0.108 & 0.332 & -0.325 & 0.747 & -0.772 & 0.556 \\ 
\multicolumn{1}{l}{\coef{lab} (lab)} & 0.413 & 0.338 & 1.224 & 0.228 & -0.262 & 1.089 \\ 
\multicolumn{1}{l}{\coef{experience} (experience)} & 0.941 & 0.361 & 2.607 & 0.013 & 0.219 & 1.663 \\ 
\multicolumn{1}{l}{\coef{ability} (ability)} & 0.863 & 0.297 & 2.904 & 0.006 & 0.269 & 1.457 \\ 
\bottomrule \\[-1.8ex] 
\end{tabular} 
\end{table*} 


\autoref{tab:mvr-effe-freq-gauss} shows the results of fitting model \eqref{eq:basic-lr}
using a standard least-squares algorithm.
Each predictor variable $\var{v}$'s coefficient $\coef{v}$ gets a maximum-likelihood \emph{estimate} $\estimate{v}$ and a \emph{standard error} $\error{v}$ of the estimate.
The numbers are a bit different from those in Table~III of \autotest even if we used their replication package;
after discussion\footnote{Personal communication between Torkar and Ceccato, December 2017.}
with the authors of \cite{autotest-bib}, we attribute any remaining differences 
to a small mismatch in how the data has been reported in the replication package.
Importantly, the difference does not affect the overall findings, and hence \autoref{tab:mvr-effe-freq-gauss} is a sound substitute of \autotest's Table~III for our purposes.

\myparagraph{Null hypothesis.}
To understand whether any of the predictor variables $v$ in \eqref{eq:basic-lr} 
makes a significant contribution to the number of \var{fixed} bugs,
it is customary to define a \emph{null hypothesis} $H_{0}^v$, 
which expresses the fact that debugging effectiveness does \emph{not} 
depend on the value of $v$.
For example, for predictor \var{treatment}:
\begin{description}
\item[$H_0^\var{treatment}$:] 
there is no difference in the effectiveness of debugging 
when debugging is supported by autogen or manual tests.
\end{description}

Under the linear model \eqref{eq:basic-lr},
if $H_0^v$ is true then regression coefficient $\coef{v}$ should be equal to zero.
However, error term $\epsilon$ makes model \eqref{eq:basic-lr} stochastic;
hence, we need to determine the \emph{probability} that some coefficients are zero.
If the probability that $\coef{v}$ is close to zero is \emph{small} (typically, less than 5\%)
we customarily say that the corresponding variable $\var{v}$ is a \emph{statistically significant predictor}
of \var{fixed}; and, correspondingly, we reject $H_0^v$.

\myparagraph{$p$-values.}
The other columns in \autoref{tab:mvr-effe-freq-gauss} 
present standard frequentist statistics used in hypothesis testing.
Column $t$ is the $t$-statistic, which is simply the coefficient's estimate 
divided by its standard error.
One could show~\cite{studentT} that, if the null hypothesis $H_0^v$ were \emph{true},
its $t$ statistic $t_v$ over repeated experiments 
would follow a $t$ distribution $t(n_o - n_\beta)$ with $n_o - n_\beta$ degrees of freedom,
where $n_o$ is the number of observations (the sample size),
and $n_\beta$ is the number of coefficients in \eqref{eq:basic-lr}.

The $p$-value is the probability of observing data 
that is, under the null hypothesis, at least as extreme as the one that was actually observed.
Thus, the $p$-value for $v$ is
the probability that $t_v$ takes a value smaller than $-|\overline{t_v}|$ or greater than $+|\overline{t_v}|$, 
where $\overline{t_v}$ is the value of $t_v$ observed in the data;
in formula:
\begin{equation}
p_v
\ =\ 
\int_{-\infty}^{-|t_v|} \dist{T}(x) \de x
+
\int^{+\infty}_{+|t_v|} \dist{T}(x) \de x
\,,
\label{eq:pval-t}
\end{equation}
where $\dist{T}$ is the density function of distribution $t(n_o - n_\beta)$.
Column $p$-value in \autoref{tab:mvr-effe-freq-gauss} reports $p$-values 
for every predictor variable computed according to \eqref{eq:pval-t}.

\subsubsection{Significance Analysis}
\label{sec:autotest:significance}

\myparagraph{Significant predictors.}
Using a standard confidence level $\alpha = 0.05$,
\autotest's analysis concludes that:
\begin{itemize}
\item \var{system} and \var{lab} are \emph{not statistically significant} 
because their $p$-values are much larger than $\alpha$---in other words, 
we cannot reject $H_0^{\var{system}}$ or $H_0^{\var{lab}}$;

\item \var{experience} and \var{ability} are \emph{statistically significant}
because their $p$-values are strictly smaller than $\alpha$---in other words,
we reject $H_0^{\var{experience}}$ and $H_0^{\var{ability}}$;

\item \var{treatment} is \emph{borderline} significant
because its $p$-value is close to $\alpha$---in other words,
we tend to reject $H_0^{\var{treatment}}$.
\end{itemize}

Unfortunately, as we discussed in \autoref{sec:bayes-overview},
$p$-values cannot reliably assess statistical significance.
The problem is that the $p$-value is a 
probability \emph{conditional on the null hypothesis being true}:
$p_v = \prob[|t_v| > |\overline{t_v}| \mid H_0^v]$;
if $p_v \ll \alpha$ is small, the data is unlikely to happen under the null hypothesis,
but this is not enough to soundly conclude whether the null hypothesis itself is likely true.

Another, more fundamental, problem is that null hypotheses are simply too restrictive, and hence unlikely to be literally true.
In our case, a coefficient may be statistically significant from zero but still very small in value;
from a practical viewpoint, this would be as if it were zero.
In other words, statistical significance based on null hypothesis testing may be irrelevant to assess \emph{practical significance}~\cite{Bayes1}.

\myparagraph{Confidence intervals.}
Confidence intervals are a better way of assessing statistical significance 
based on the output of a fitted linear regression model.
The most common 95\% confidence interval is based on the normal distribution, i.e.,
\[
\ci{v}
\ =\ 
(
\estimate{v} - 2\, \error{v},\ 
\estimate{v} + 2\, \error{v}
)
\]
is the 95\% confidence interval for a variable \var{v} whose coefficient 
\coef{v} has estimate \estimate{v} and standard error \error{v} given by the linear regression fit.

If confidence interval $\ci{v}$ includes zero,
it means that there is a significant chance that $\coef{v}$ is zero;
hence, we conclude that $v$ is not significant with 95\% confidence.
Conversely, if $\ci{v}$ for variable $v$ does not include zero,
it includes either only positive or only negative values;
hence, there is a significant chance that $\coef{v}$ is not zero, and
we conclude that $v$ is significant with 95\% confidence.

Even though, with \autotest's data, 
it gives the same qualitative results as the analysis based on $p$-values,
grounding significance on confidence intervals has the distinct advantage of
providing a \emph{range of values} the coefficients of a linear regression model 
may take based on the data---instead of the $p$-value's conditional probability that 
really answers not quite the right question.

\myparagraph{Interpreting confidence intervals.}
It is tempting to interpret a frequentist confidence intervals $\ci{v}$ 
in terms of probability of the possible values of the regression coefficient $\coef{v}$ of variable $v$.\footnote{This interpretation is widespread even though it is incorrect~\cite{robustMisinterpretation}.}
Nonetheless the correct frequentist interpretation is much less serviceable~\cite{MoreyHRLW2016CI}.
First, the 95\% probability does not refer to the values of $\coef{v}$ relative to the specific interval $\ci{v}$ we computed,
but rather to the \emph{process} used to compute the confidence interval:
in an infinite series of repeated experiments---in each of which we compute a different interval $\ci{v}$---the
fraction of repetitions where the \emph{true} value of $\coef{v}$ falls in $\ci{v}$ will tend to 95\% in the limit.
Second, confidence intervals have no distributional information: ``there is no direct sense by which parameter values in the middle of the confidence interval are more probable than values at the ends of the confidence interval''~\cite{bayesianNewStats-ROPE}.

To be able to soundly interpret $\ci{v}$ in terms of probabilities we need to assign a \emph{prior} probability to the regression coefficients and apply Bayes's theorem.
A natural assumption is to use a uniform prior; then, under the normal model for the error term $\epsilon$ in \eqref{eq:basic-lr-0},
$v$ behaves like a random variable with a normal distribution with mean equal to $v$'s linear regression estimate \estimate{v} and standard deviation equal to $v$'s standard error \error{v}.
Therefore, $\ci{v} = (\cilow, \cihigh)$ is the (centered) interval such that
\[
\int_{\cilow}^{\cihigh} \dist{N}(x) \de x
\ =\ 
0.95\,,
\]
where $\dist{N}$ is the normal density function of distribution
$\normal{\estimate{v}}{\error{v}}$---with mean equal to $\coef{v}$'s estimate 
and standard deviation equal to $\coef{v}$'s standard error.

However, this entails that if $\ci{v}$ lies to the right of zero ($\cilow > 0$)
then the probability that the true value of $\coef{v}$ is positive is
\begin{align*}
  \int_{0}^{+\infty} \dist{N}(x)\de x
  &\ \geq\ 
    \int_{\cilow}^{+\infty} \dist{N}(x)\de x
  \\
  &\ =\ 
  \int_{\cilow}^{\cihigh} \dist{N}(x) \de x
  +
  \int_{\cihigh}^{+\infty} \dist{N}(x) \de x
  \\
  &\ =\
  0.95 + 0.025\,,
\end{align*}
that is at least 97.5\%.

\subsubsection{Summary and Criticism of Frequentist Statistics}
\label{sec:autotest:summary-freq}

In general, assessing statistical significance using $p$-values is hard to justify methodologically, since the conditional probability that $p$-values measure is neither sufficient nor necessary to decide whether the data supports or contradicts the null hypothesis.
Confidence intervals might be a better way to assess significance; however, they are surprisingly tricky to interpret in a purely frequentist setting as perusing \autotest's analysis has shown. More precisely, we have seen that any sensible, intuitive interpretation of confidence intervals requires the introduction of \emph{priors}.
We might as well go all in and perform a full-fledged Bayesian analysis instead;
as we will demonstrate in \autoref{sec:bayesian-testing}, this generally brings greater flexibility and other advantages.

\subsection{Bayesian Reanalysis}
\label{sec:bayesian-testing}

\subsubsection{Linear Regression Model}
\label{sec:autotest:bayes-linear-model}

As a first step, let us apply Bayesian techniques to fit the very same linear model \eqref{eq:basic-lr} used in \autotest's analysis.
\autoref{tab:mvr-effe-bayes-gauss} summarizes the Bayesian fit with statistics similar to those used in the frequentist setting: for each predictor variable $\var{v}$, we estimate the value and standard error of multiplicative coefficient $\coef{v}$---as well as the endpoints of the 95\% credible interval (last two columns in \autoref{tab:mvr-effe-bayes-gauss}).

\begin{table*}[!htbp] \centering 
  \caption{Regression coefficients in the linear model with Gaussian error model \eqref{eq:basic-lr}, computed with Bayesian analysis.
    For each coefficient \coef{v}, the table reports
    posterior mean \emph{estimate} $\estimate{v}$,
    posterior standard deviation (or \emph{error}) \error{v} of the estimate,
    and \emph{lower} and \emph{upper} endpoints of the 95\% uncertainty interval of the coefficient 
    (corresponding to 2.5\% and 97.5\% cumulative posterior probability).} 
  \label{tab:mvr-effe-bayes-gauss} 
  \normalsize
\begin{tabular}{@{\extracolsep{5pt}} D{.}{.}{-3} D{.}{.}{-3} D{.}{.}{-3} D{.}{.}{-3} D{.}{.}{-3} } 
\toprule \\ [-2ex] 
\multicolumn{1}{c}{\textsc{coefficient} \coef{v}} & \multicolumn{1}{c}{\textsc{estimate} $\estimate{v}$} & \multicolumn{1}{c}{\textsc{error} \error{v}} & \multicolumn{1}{c}{\textsc{lower}} & \multicolumn{1}{c}{\textsc{upper}} \\ 
\midrule \\[-1.8ex] 
\multicolumn{1}{l}{\coef{intercept} (intercept)} & -1.753 & 0.719 & -3.159 & -0.319 \\ 
\multicolumn{1}{l}{\coef{treatment} (treatment)} & 0.659 & 0.337 & 0.011 & 1.332 \\ 
\multicolumn{1}{l}{\coef{system} (system)} & -0.112 & 0.347 & -0.808 & 0.566 \\ 
\multicolumn{1}{l}{\coef{lab} (lab)} & 0.408 & 0.344 & -0.265 & 1.069 \\ 
\multicolumn{1}{l}{\coef{experience} (experience)} & 0.947 & 0.363 & 0.245 & 1.670 \\ 
\multicolumn{1}{l}{\coef{ability} (ability)} & 0.863 & 0.298 & 0.290 & 1.443 \\ 
\bottomrule \\[-1.8ex] 
\end{tabular} 
\end{table*}

The 95\% \emph{uncertainty interval} for variable $\var{v}$ is the centered\footnote{That is, centered around the distribution's mean.} interval $(\ell, h)$ such that \[
\int_{\cilow}^{\cihigh} \dist{P}_\var{v}(x) \de x
\ =\ 
0.95\,,
\]
where $\dist{P}_\var{v}$ is the posterior distribution of coefficient $\coef{v}$.
Since $\dist{P}_\var{v}$ is the actual distribution observed on the data, the uncertainty interval can be interpreted in a natural way as the range of values such that $\coef{v} \in (\ell, h)$ with probability 95\%.

Uncertainty intervals are traditionally called \emph{credible intervals}, but we agree with Gelman's suggestion\footnote{Andrew Gelman's blog \href{https://andrewgelman.com/2010/12/21/lets_say_uncert/}{\emph{Instead of ``confidence interval,'' let’s say ``uncertainty interval''}}, 21 December 2010}
that \emph{uncertainty interval} reflects more clearly what they measure.\footnote{More recently, other statisticians suggested yet another term: \emph{compatibility interval}~\cite{compatibility}.}
We could use the same name for the frequentist confidence intervals, but we prefer to use different terms to be able to distinguish them easily in writing.

The numbers in \autoref{tab:mvr-effe-bayes-gauss} summarize a multivariate distribution on the regression coefficients.
The distribution---called the \emph{posterior distribution}---is obtained, as described in \autoref{sec:bayesian-inference-algo},
by repeatedly sampling values 
from a \emph{prior} distribution, and applying Bayes' theorem using a likelihood function derived from \eqref{eq:basic-lr} (in other words, based on the normal error term).
The posterior is thus a sampled, numerical distribution, which provides rich information about the possible range of values for the coefficients, and lends itself to all sorts of derived computations---as we will demonstrate in later parts of the reanalysis.

\myparagraph{Statistical significance.}
Using the same conventional threshold used by the frequentist analysis, we stipulate that a predictor \var{v} is \emph{statistically significant} if the 95\% uncertainty interval of its regression coefficient $\coef{v}$ does not include zero. According to this definition, \var{treatment}, \var{experience}, and \var{ability} are all significant---with \var{treatment} more weakly so.
While the precise estimates differ, the big picture is consistent with the frequentist analysis---but this time the results' interpretation is unproblematic.

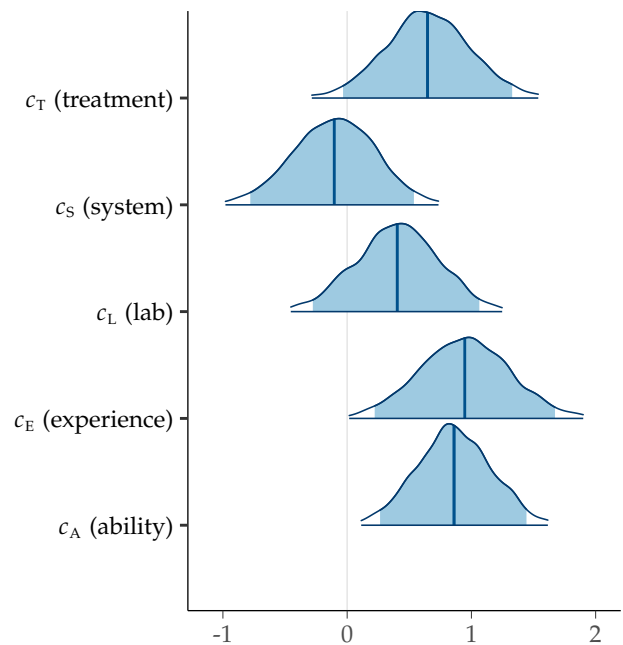
\begin{figure}[!hbt]
  \begin{tikzpicture}\end{tikzpicture}
  \caption{Plots of the posterior probability distributions of each regression coefficient in the linear model with Gaussian error model \eqref{eq:basic-lr}, computed with Bayesian analysis and weak unbiased priors.
    Each plot is a density plot that covers an area corresponding to 99\% probability; the inner shaded area corresponds to 95\% probability; and the thick vertical line marks the distribution's median.}
  \label{fig:mvr-effe-bayes-gauss}
\end{figure}

Since the coefficients' distributions are fully available we can even plot them.
In \autoref{fig:mvr-effe-bayes-gauss}, the thick vertical lines mark the estimate (the distribution's median); the shaded areas correspond to 95\% probability; and the distribution curves extend to cover the 99\% probability overall.
Visualization supports a more nuanced view of statistical significance, because the plots include probability distribution information:\footnote{Remember that confidence intervals do not possess this kinds of information (see \autoref{sec:autotest:significance}).}
it is clear that \var{experience} and \var{ability} are strongly significant;
even though \var{treatment} is borderline significant, there remains a high probability that it is significant (since most of the curve is to the right of the origin);
while neither \var{system} nor \var{lab} is significant at the 95\% value,
we can discern a weak tendency towards significance for \var{lab}---whereas
\var{system} has no consistent effect.

\subsubsection{Generalized Models}
\label{sec:autotest:bayes-priors}
\myparagraph{Priors.}
The posterior distributions look approximately normal; 
indeed, we have already discussed how they are exactly normal
if we take uniform priors.
A uniform prior assumes that any possible value for a coefficient is as likely as any other.
In some restricted conditions (see \autoref{sec:autotest:summary-bayes}),
and with completely uniform priors, frequentist and Bayesian analysis tend to lead
to the same overall numerical results; 
however, Bayesian analysis is much more flexible because it can compute posterior distributions
with priors that are not completely uniform. 

A \emph{weak unbiased} prior is the most appropriate choice in most cases.
Even when we do not know much about the possible values that the regression coefficients might take,
we often have a rough idea of their variability range.
Take variable \var{treatment}: discovering whether it is definitely positive or negative is one of the main goals of the analysis; but we can exclude that \var{treatment} (indeed, any variable) has a huge effect on the number of fixed bugs (such as hundred or even thousands of bugs difference).
Since such a huge effect is ruled out by experience and common sense,
we used a normal distribution with mean $0$ and standard deviation $20$ as prior.
This means that we still have no bias on whether \var{treatment} is significant (mean $0$),
but we posit that its value is most likely between $-60 = 0 - 3 \cdot 20$ and $+60 = 0 + 3 \cdot 20$ (three standard deviations off the mean).\footnote{Since a normal distribution has infinitely long tails, using it as prior does not rule out any posterior value completely, but makes extreme values exceedingly unlikely without exceptionally strong data to support them (i.e., we make infinity a non-option).}
In this case, using a completely uniform prior would not lead to qualitatively different results, but would make the estimate a bit less precise without adding any relevant information.

If we had evidence---from similar empirical studies or other sources of in\-for\-ma\-tion---that sharpen our initial estimate of an effect, we could use it to pick a more biased prior.
We will explore this direction more fully in \autoref{sec:bayesian-rosetta} where we present the other case study.

\myparagraph{Poisson data model.}
To further demonstrate how Bayesian modeling is flexible and supports quantitative predictions,
let us modify \eqref{eq:basic-lr} to build a \emph{generalized} linear model of the \autotest's experiments.

Since \var{system} and \var{lab} turned out to be not significant,
we exclude them from the new model.
Excluding insignificant variables has the advantage of simplifying the model
without losing accuracy---thus supporting better generalizations~\cite{rethinking}.
Then, since \var{fixed} can only be a nonnegative integer, we model it as drawn from a Poisson distribution (suitable for counting variables) rather than a normal distribution as in \eqref{eq:basic-lr} (which allows for negative real values as well).
Finally, we use a stronger prior for \coef{treatment} corresponding to a normal distribution $\dist{N}(0.5, 0.8)$,
which is a bit more biased as it nudges \coef{treatment} towards positive values of smallish size:
\begin{equation}
\var{fixed}
\sim  
\dist{P}\left(
  \exp\left(
  \begin{array}{rl}
   \coef{intercept} &\!\!\!+\ \coef{treatment}\cdot\var{treatment} \\
                    &\!\!\!+\ \coef{experience}\cdot \var{experience} \\
                    &\!\!\!+\ \coef{ability}\cdot \var{ability}
  \end{array}
\right)\right)
\label{eq:poisson-glm}
\end{equation}
where $\dist{P}(\lambda)$ is the Poisson distribution with rate $\lambda$.
Since $\lambda$ has to be positive, it is customary to take the \emph{exponential}
of the linear combination of predictor variables as a parameter of the Poisson.

\begin{table*}[!htbp] \centering 
  \caption{Regression coefficients in the generalized linear model with Poisson error model \eqref{eq:poisson-glm}, computed with Bayesian analysis.
    For each coefficient \coef{v}, the table reports
    posterior mean \emph{estimate} $\estimate{v}$,
    posterior standard deviation (or \emph{error}) \error{v} of the estimate,
    and \emph{lower} and \emph{upper} endpoints of the 95\% uncertainty interval of the coefficient 
    (corresponding to 2.5\% and 97.5\% cumulative posterior probability).} 
  \label{tab:mvr-effe-bayes-poisson-reduced}
  \normalsize
\begin{tabular}{@{\extracolsep{5pt}} D{.}{.}{-3} D{.}{.}{-3} D{.}{.}{-3} D{.}{.}{-3} D{.}{.}{-3} } 
\toprule \\ [-2ex]
\multicolumn{1}{c}{\textsc{coefficient} \coef{v}} & \multicolumn{1}{c}{\textsc{estimate} $\estimate{v}$} & \multicolumn{1}{c}{\textsc{error} \error{v}} & \multicolumn{1}{c}{\textsc{lower}} & \multicolumn{1}{c}{\textsc{upper}} \\ 
\midrule \\[-1.8ex] 
\multicolumn{1}{l}{\coef{intercept} (intercept)} & -1.953 & 0.510 & -3.006 & -0.954 \\ 
\multicolumn{1}{l}{\coef{treatment} (treatment)} & 0.493 & 0.254 & 0.012 & 0.991 \\ 
\multicolumn{1}{l}{\coef{experience} (experience)} & 0.800 & 0.309 & 0.201 & 1.400 \\ 
\multicolumn{1}{l}{\coef{ability} (ability)} & 0.642 & 0.226 & 0.205 & 1.096 \\ 
\bottomrule \\[-1.8ex] 
\end{tabular} 
\end{table*}

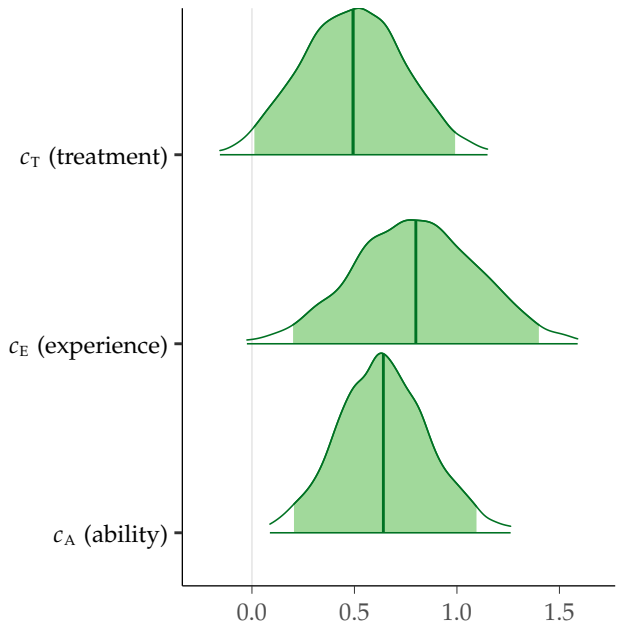
\begin{figure}[!hbt]
  \centering
  \begin{tikzpicture}\end{tikzpicture}
  \caption{Plots of the posterior probability distributions of each regression coefficient in the generalized linear model with Poisson error model \eqref{eq:poisson-glm}, computed with Bayesian analysis.
    Each plot is a density plot that covers an area corresponding to 99\% probability; the inner shaded area corresponds to 95\% probability; and the thick vertical line marks the distribution's median.}
  \label{fig:mvr-effe-poisson-reduced}
\end{figure}

Fitting model \eqref{eq:poisson-glm} gives the estimates in \autoref{tab:mvr-effe-bayes-poisson-reduced}, graphically displayed in \autoref{fig:mvr-effe-poisson-reduced}.
The results are still qualitatively similar to previous analyses: \var{treatment}, \var{experience}, and \var{ability} are significant at the 95\% level---\var{treatment} more weakly so.
However, a simpler model is likely, all else being equal, to perform better predictions;
let us demonstrate this feature in the next section.

\subsubsection{Quantitative Prediction}
\label{sec:autotest:quantitative}
Simplicity is a definite advantage of linear regression models:
simple models are easy to fit and, most important, easy to interpret.
After estimating the coefficients in \eqref{eq:basic-lr-0},
we can use those estimates to perform quantitative predictions.
For example, in the linear model \eqref{eq:basic-lr-0},
$\estimate{treatment} = 0.66$ is the average difference
in the number of bugs that a programmer working with autogen tests could fix
compared to working with manual tests---all other factors staying the same:
\[
\begin{split}
  & \left(
    \begin{array}{rl}
      \estimate{intercept}
      & + \overbrace{\estimate{treatment} \cdot 1}^{\var{treatment}\,=\,\text{auto}}
        +\: \estimate{system} \cdot \var{system} + \estimate{lab} \cdot \var{lab}
      \\
      &  +\: \estimate{experience} \cdot \var{experience} + \estimate{ability} \cdot \var{ability}
        + \epsilon
    \end{array}
  \right)
\\
- &
\left(
    \begin{array}{rl}
      \estimate{intercept}
      & + \overbrace{\estimate{treatment} \cdot 0}^{\var{treatment}\,=\,\text{manual}}
        +\: \estimate{system} \cdot \var{system}
        + \estimate{lab} \cdot \var{lab}
      \\
      & +\: \estimate{experience} \cdot \var{experience}
        + \estimate{ability} \cdot \var{ability}
        + \epsilon
    \end{array}
  \right)
\\
= &
\quad
\estimate{treatment}
\end{split}
\]

Even though the transparency of the linear model is a clear plus,
performing estimates and predictions only based on the \emph{estimates}
of the regression coefficients has its limits.
On the one hand, we ignore the richer information---in the form of posterior distributions---that
Bayesian analyses provide.
On the other hand, analytic predictions on the fitted model may become cumbersome
on more complex models, such as the Poisson regression model \eqref{eq:poisson-glm},
where the information carried by regression coefficients is less intuitively understandable.
Ideally, we would like to select a model based on how well it can characterize the data
without having to trade this off against mainly computational issues.

Since Bayesian techniques are based on numerical methods,
they shine at estimating and predicting derived statistics.
We can straightforwardly \emph{sample} the predictor input space in a way
that reflects the situation under study, and \emph{simulate} the outcomes numerically.\footnote{Since frequentist statistics
  do not have post-data distributional information, this can be done only within a Bayesian framework.}
Provided the sample is sufficiently large, the average outcome reflects the fitted model's predictions.
Let us show one such simulation to try to answer a relevant and practical software engineering question based on the data at hand. 

Our analysis has shown that autogen tests are better---lead to more fixed bugs---than manual tests;
and that programmers with higher ability are also better at fixing bugs.
Since high-ability programmers are a clearly more expensive resource than automatically generated tests, can we use autogen tests extensively to partially make up for the lack of high-ability programmers?
We instantiate the question in quantitative terms:

\begin{quote}
  How does the bug-fixing performance of these two teams of programmers compare?
  \begin{description}
  \item[Team low\slash auto:] made of 80\% low-ability, 10\% medium-ability, and 10\%~high-ability programmers;
                 using 10\% manual tests and 90\% autogen tests
  \item[Team high\slash manual:] made of 40\% low-ability, 40\% medium-ability, and 20\%~high-ability programmers;
                 using 50\% manual tests and 50\% autogen tests
  \end{description}
\end{quote}

Simulating the performance of these two teams using model \eqref{eq:poisson-glm} fitted
as in \autoref{tab:mvr-effe-bayes-poisson-reduced},
we get that team high\slash manual fixes 20\% more bugs than team low\slash auto in the same conditions.
This analysis tells us that ability is considerably more decisive than the nature of tests, but autogen tests
can provide significant help for a modest cost.
It seems likely that industrial practitioners could be interested in simulating several such scenarios to get evidence-based support for decisions and improvement efforts. In fact, the posterior distributions obtained from a Bayesian statistical analysis could be used to optimize, by repeated simulation, different variables---the number of high-ability team members and the proportion of manual versus autogen test cases in our example---to reach specific goals identified as important by practitioners.

\subsubsection{Summary of Bayesian Statistics}
\label{sec:autotest:summary-bayes}
Using Bayesian techniques brings advantages over using frequentist techniques even when the two approaches lead to the same qualitative results.
Bayesian models have an unambiguous probabilistic interpretation,
which relates the variables of interest to their probabilities of taking certain values.
A fitted model consists of \emph{probability distributions} of the estimated parameters,
rather than less informative point estimates.
In turn, this supports quantitative estimates and predictions of arbitrary scenarios based on numerical simulations, and fosters nuanced data analyses that are grounded in practically significant measures.

Numerical simulations are also possible in a frequentist setting,
but their soundness often depends on a number of subtle assumptions that are easy to overlook or misinterpret.
In the best case, frequentist statistics is just an inflexible
hard-to-interpret version of Bayesian statistics; in the worst case,
it may even lead to unsound conclusions.

\myparagraph{Frequentist $\neq$ Bayesian $+$ Uniform Priors.}
\label{sec:no-correspondence-theorem}
In this case study, the Bayesian analysis's main results turned out to be numerically quite close to the frequentist's.
We also encountered several cases where
the ``intuitive'' interpretation of measures such as confidence intervals is incorrect under a strict frequentist interpretation
but can be justified by assuming a Bayesian interpretation and uniform (or otherwise ``uninformative'') priors.
It is tempting to generalize this and think that frequentist statistics are mostly Bayesian statistics with uniform priors by other means.
Unfortunately, this is in general \emph{not} true:
on the contrary, the correspondence between frequentist statistics and Bayesian statistics with uninformative priors holds only in few simple cases.

Specifically, frequentist confidence intervals and Bayesian uncertainty intervals calculations
coincide only in the simple case of estimating the mean of a normal distribution---what we did
with the  simple linear regression model \eqref{eq:basic-lr}---when
the Bayesian calculation uses a particular uninformative prior~\cite{jeffreys-book}.
In many other applications~\cite{MoreyHRLW2016CI}, frequentist confidence intervals markedly differ from Bayesian uncertainty intervals,
and the procedures to calculate the former do not normally come with any guarantees that they can be validly interpreted as post-data statistics.

\section{Direct Bayesian Modeling: The Role of Priors}
\label{sec:rosetta}

Comparing programming languages to establish which is better at a
certain task is an evergreen topic in computer science research;
empirical studies have made such comparisons more rigorous
and their outcomes more general.

In this line of work, \rosetta analyzed programs in the Rosetta Code wiki~\cite{rosettacode}.
Rosetta Code is a collection of programming tasks, each implemented in various programming languages.
The variety of tasks, and the expertise and number of contributors to the Rosetta Code wiki,
buttresses a comparison of programming languages under conditions that are more natural than few performance benchmarks,
yet still more controlled than empirical studies of large code repositories in the wild.

\subsection{Data: Performance of Programming Languages}
\label{sec:data-rosetta}

\rosetta compare eight programming languages for features such as conciseness and performance, based on experiments with a curated selection of programs from the Rosetta Code repository~\cite{rosettacode}.
Our renalysis focuses on \rosetta's analysis of running time \emph{performance}.

For each language $\ell$ among C, C\#, F\#, Go, Haskell, Java, Python, and Ruby, \rosetta's experiments involve an ordered set $T(\ell)$ of programming \emph{tasks}, such as sorting algorithms, combinatorial puzzles, and NP-complete problems; each task comes with an explicit sample input.
For a task $t \in T(\ell)$, $R(\ell, t)$ denotes the set of running times of any implementations of task $t$ in language $\ell$, among those available in Rosetta Code, that ran without errors or timeout on the same sample input.
$R(\ell, t)$ is a \emph{set} because Rosetta Code often includes several implementation of the same task (algorithm) in the same language---variants
that may explore different programming styles or were written by different contributors.

Take a \emph{pair} $\ell_1, \ell_2$ of languages; $T_{1,2}$ is the set $T(\ell_1) \cap T(\ell_2)$ of tasks that have implementations in both languages.
Thus, we can directly compare the performance of $\ell_1$ and $\ell_2$ on any task $t \in T_{1,2}$ by juxtaposing each running time in $R(\ell_1, t)$ to each in $R(\ell_2, t)$.
To this end, define $R_{1,2}(t)$ as the Cartesian product $R(\ell_1, t) \times R(\ell_2, t)$: a component $(r_1, r_2)$ of $R_{1,2}(t)$ measures the running time $r_1$ of some implementation of a task in language $\ell_1$ against the running time $r_2$ of some implementation of the same task in the other language $\ell_2$.

It is useful to summarize each such comparison in terms of the \emph{speedup} of one implementation in one language relative to one in the other language.
To this end, $S_{1,2}(t)$ is a vector of the same length as $R_{1,2}(t)$, such that each component $(r_1, r_2)$ in the latter determines component $\iota(r_1, r_2)$ in the former, where 
\begin{equation}
\label{eq:ratio-def}
\iota(a, b)\ =\ \sgn(a - b) \cdot \left(1 - \frac{\min(a, b)}{\max(a, b)} \right).
\end{equation}
The signed ratio $\iota(r_1, r_2)$ varies over the interval $(-1, +1)$:
it is zero iff $r_1 = r_2$ (the two implementations ran in the same time exactly),
it is negative iff $r_1 < r_2$ (the implementation in language $\ell_1$ was faster than the implementation in language $\ell_2$), and it is positive iff $r_1 > r_2$ (the implementation in language $\ell_2$ was faster than the implementation in language $\ell_1$).
The absolute value of the signed ratio is proportional to how much faster the faster language is relative to the slower language; precisely, $1/(1 - \left|\iota(r_1, r_2)\right|)$ is the faster-to-slower speedup ratio.
\rosetta directly compared such speedup ratios, but here we prefer to use the \emph{inverse speedups} $\iota(r_1, r_2)$ because they encode the same information but using a smooth function that ranges over a bounded interval, which is easier to model and to compare to statistics, such as effect sizes, that are also normalized.

Among the various implementations of a chosen task in each language, it makes sense to focus on the \emph{best} one relative to the measured features; that is, to consider the \emph{fastest} variant.
We use identifiers with a hat to denote the fastest measures, corresponding to the smallest running times:
$\widehat{R}(\ell, t)$ is thus $\min R(\ell, t)$, and $\widehat{R}_{1,2}(t)$ and $\widehat{S}_{1,2}(t)$ are defined like their unhatted counterparts but refer to $\widehat{R}(\ell_1, t)$ and $\widehat{R}(\ell_2, t)$.
In the following, we refer to the hatted data as the \emph{optimal data}.

\myparagraph{Analysis goals.}
The main goal of \rosetta's analysis is to determine which languages
are faster and which are slower based on the empirical performance data described above.
The performance comparison is pairwise:
for each language pair $\ell_1, \ell_2$,
we want to determine:

\begin{enumerate}
\item whether the performance difference between $\ell_1$ and $\ell_2$ is significant;
\item if it is significant, how much $\ell_1$ is faster than $\ell_2$.
\end{enumerate}

\subsection{Frequentist Analysis}
\label{sec:frequentist-rosetta}

\autoref{sec:rosetta-pairwise} follows closely \rosetta's analysis using null hypothesis testing and effect sizes.
Then, \autoref{sec:rosetta-multiple} refines the frequentist analysis by taking into account the multiple comparisons problem~\cite{Miller-multistats}.

\subsubsection{Pairwise Comparisons}
\label{sec:rosetta-pairwise}

\myparagraph{Null hypothesis.}
As customary in statistical hypothesis testing,
\rosetta use null hypotheses to express whether differences are significant or not.
For each language pair $\ell_1, \ell_2$, null hypothesis $H_0^{1,2}$ denotes lack of significant difference between the two languages:
\begin{description}
\item[$H_0^{1,2}$:] 
  there is no difference in the performance of languages $\ell_1$ and $\ell_2$
  when used for the same computational tasks
\end{description}

\rosetta use the Wilcoxon signed-rank test to compute the $p$-value
expressing the probability of observing performance data
($\widehat{R}(\ell_1, t)$ against $\widehat{R}(\ell_2, t)$ for every $t \in T_{1,2}$)
at least as extreme as the one observed, assuming the null hypothesis $H_0^{1,2}$.
The Wilcoxon signed-rank test was chosen because
it is a paired, unstandardized test (it does not require normality of the data).
Thus, given a probability $\alpha$,
if $p < \alpha$ we reject the null hypothesis and say that the difference between $\ell_1$ and $\ell_2$ is \emph{statistically significant at level~$\alpha$}.

\myparagraph{Effect size.}
Whenever the analysis based on $p$-values indicates that the difference
between $\ell_1$ and $\ell_2$ is significant,
we still do not know whether $\ell_1$ or $\ell_2$ is the faster language of the two,
nor how much faster it is.
To address this question, we compute Cliff's $\delta$ effect size,
which is a measure of how often the measures of one language are smaller than the measures of the other language;\footnote{\rosetta's analysis used Cohen's $d$ as effect size; here we switch to Cliff's $\delta$ both because it is more appropriate for nonparametric data and because it varies between $-1$ and $+1$ like the \emph{inverse} speedup ratio $\iota$ we use in the reanalysis.}
precisely, $\delta < 0$ means that $\ell_1$ is faster than $\ell_2$ on average---the closer $\delta$ is to $-1$ the higher $\ell_1$'s relative speed;
conversely, $\delta > 0$ means that $\ell_2$ is faster than $\ell_1$ on average.

\begin{table*}[!h]
  \setlength{\tabcolsep}{5pt}  
  \centering
  \footnotesize
  \begin{tabular}{llccccccc}
\toprule
\textsc{language} & \textsc{measure} & \multicolumn{1}{r}{\color{lang1}{C}} & \multicolumn{1}{r}{\color{lang1}{C\#}} & \multicolumn{1}{r}{\color{lang1}{F\#}} & \multicolumn{1}{r}{\color{lang1}{Go}} & \multicolumn{1}{r}{\color{lang1}{Haskell}} & \multicolumn{1}{r}{\color{lang1}{Java}} & \multicolumn{1}{r}{\color{lang1}{Python}} \\ 
\midrule
\multicolumn{1}{r}{\nopagebreak \color{lang2}{C\#}} & \multicolumn{1}{r}{\nopagebreak $p$}  & \multicolumn{1}{r}{\color{sstrong}{$0.000$}} & \multicolumn{1}{r}{} & \multicolumn{1}{r}{} & \multicolumn{1}{r}{} & \multicolumn{1}{r}{} & \multicolumn{1}{r}{} & \multicolumn{1}{r}{} \\
 & \multicolumn{1}{r}{\nopagebreak $\delta$}  & \multicolumn{1}{r}{\color{lang1}{$-0.440$}} & \multicolumn{1}{r}{} & \multicolumn{1}{r}{} & \multicolumn{1}{r}{} & \multicolumn{1}{r}{} & \multicolumn{1}{r}{} & \multicolumn{1}{r}{} \\
 & \multicolumn{1}{r}{\nopagebreak $m$}  & \multicolumn{1}{r}{$-0.793$} & \multicolumn{1}{r}{} & \multicolumn{1}{r}{} & \multicolumn{1}{r}{} & \multicolumn{1}{r}{} & \multicolumn{1}{r}{} & \multicolumn{1}{r}{} \\
 & \multicolumn{1}{r}{\nopagebreak $\mu$}  & \multicolumn{1}{r}{$-0.600$} & \multicolumn{1}{r}{} & \multicolumn{1}{r}{} & \multicolumn{1}{r}{} & \multicolumn{1}{r}{} & \multicolumn{1}{r}{} & \multicolumn{1}{r}{} \\
\multicolumn{1}{r}{\rule{0pt}{1.1\normalbaselineskip}\color{lang2}{F\#}} & \multicolumn{1}{r}{\nopagebreak $p$}  & \multicolumn{1}{r}{\color{smedium}{$0.013$}} & \multicolumn{1}{r}{$0.182$} & \multicolumn{1}{r}{} & \multicolumn{1}{r}{} & \multicolumn{1}{r}{} & \multicolumn{1}{r}{} & \multicolumn{1}{r}{} \\
 & \multicolumn{1}{r}{\nopagebreak $\delta$}  & \multicolumn{1}{r}{\color{lang1}{$-0.603$}} & \multicolumn{1}{r}{\color{lang1}{$-0.223$}} & \multicolumn{1}{r}{} & \multicolumn{1}{r}{} & \multicolumn{1}{r}{} & \multicolumn{1}{r}{} & \multicolumn{1}{r}{} \\
 & \multicolumn{1}{r}{\nopagebreak $m$}  & \multicolumn{1}{r}{$-0.875$} & \multicolumn{1}{r}{$-0.429$} & \multicolumn{1}{r}{} & \multicolumn{1}{r}{} & \multicolumn{1}{r}{} & \multicolumn{1}{r}{} & \multicolumn{1}{r}{} \\
 & \multicolumn{1}{r}{\nopagebreak $\mu$}  & \multicolumn{1}{r}{$-0.693$} & \multicolumn{1}{r}{$-0.370$} & \multicolumn{1}{r}{} & \multicolumn{1}{r}{} & \multicolumn{1}{r}{} & \multicolumn{1}{r}{} & \multicolumn{1}{r}{} \\
\multicolumn{1}{r}{\rule{0pt}{1.1\normalbaselineskip}\color{lang2}{Go}} & \multicolumn{1}{r}{\nopagebreak $p$}  & \multicolumn{1}{r}{\color{sstrong}{$0.000$}} & \multicolumn{1}{r}{\color{sstrong}{$0.006$}} & \multicolumn{1}{r}{\color{smedium}{$0.030$}} & \multicolumn{1}{r}{} & \multicolumn{1}{r}{} & \multicolumn{1}{r}{} & \multicolumn{1}{r}{} \\
 & \multicolumn{1}{r}{\nopagebreak $\delta$}  & \multicolumn{1}{r}{\color{lang1}{$-0.298$}} & \multicolumn{1}{r}{\color{lang2}{$0.270$}} & \multicolumn{1}{r}{\color{lang2}{$0.408$}} & \multicolumn{1}{r}{} & \multicolumn{1}{r}{} & \multicolumn{1}{r}{} & \multicolumn{1}{r}{} \\
 & \multicolumn{1}{r}{\nopagebreak $m$}  & \multicolumn{1}{r}{$-0.387$} & \multicolumn{1}{r}{$0.717$} & \multicolumn{1}{r}{$0.727$} & \multicolumn{1}{r}{} & \multicolumn{1}{r}{} & \multicolumn{1}{r}{} & \multicolumn{1}{r}{} \\
 & \multicolumn{1}{r}{\nopagebreak $\mu$}  & \multicolumn{1}{r}{$-0.474$} & \multicolumn{1}{r}{$0.384$} & \multicolumn{1}{r}{$0.534$} & \multicolumn{1}{r}{} & \multicolumn{1}{r}{} & \multicolumn{1}{r}{} & \multicolumn{1}{r}{} \\
\multicolumn{1}{r}{\rule{0pt}{1.1\normalbaselineskip}\color{lang2}{Haskell}} & \multicolumn{1}{r}{\nopagebreak $p$}  & \multicolumn{1}{r}{\color{sstrong}{$0.000$}} & \multicolumn{1}{r}{$0.142$} & \multicolumn{1}{r}{$0.929$} & \multicolumn{1}{r}{\color{smedium}{$0.025$}} & \multicolumn{1}{r}{} & \multicolumn{1}{r}{} & \multicolumn{1}{r}{} \\
 & \multicolumn{1}{r}{\nopagebreak $\delta$}  & \multicolumn{1}{r}{\color{lang1}{$-0.591$}} & \multicolumn{1}{r}{\color{lang1}{$-0.186$}} & \multicolumn{1}{r}{\color{lang2}{$0.124$}} & \multicolumn{1}{r}{\color{lang1}{$-0.356$}} & \multicolumn{1}{r}{} & \multicolumn{1}{r}{} & \multicolumn{1}{r}{} \\
 & \multicolumn{1}{r}{\nopagebreak $m$}  & \multicolumn{1}{r}{$-0.964$} & \multicolumn{1}{r}{$-0.655$} & \multicolumn{1}{r}{$-0.401$} & \multicolumn{1}{r}{$-0.909$} & \multicolumn{1}{r}{} & \multicolumn{1}{r}{} & \multicolumn{1}{r}{} \\
 & \multicolumn{1}{r}{\nopagebreak $\mu$}  & \multicolumn{1}{r}{$-0.688$} & \multicolumn{1}{r}{$-0.315$} & \multicolumn{1}{r}{$0.003$} & \multicolumn{1}{r}{$-0.440$} & \multicolumn{1}{r}{} & \multicolumn{1}{r}{} & \multicolumn{1}{r}{} \\
\multicolumn{1}{r}{\rule{0pt}{1.1\normalbaselineskip}\color{lang2}{Java}} & \multicolumn{1}{r}{\nopagebreak $p$}  & \multicolumn{1}{r}{\color{sstrong}{$0.000$}} & \multicolumn{1}{r}{$0.140$} & \multicolumn{1}{r}{\color{sstrong}{$0.008$}} & \multicolumn{1}{r}{$0.451$} & \multicolumn{1}{r}{$0.064$} & \multicolumn{1}{r}{} & \multicolumn{1}{r}{} \\
 & \multicolumn{1}{r}{\nopagebreak $\delta$}  & \multicolumn{1}{r}{\color{lang1}{$-0.446$}} & \multicolumn{1}{r}{\color{lang2}{$0.062$}} & \multicolumn{1}{r}{\color{lang2}{$0.434$}} & \multicolumn{1}{r}{\color{lang1}{$-0.147$}} & \multicolumn{1}{r}{\color{lang2}{$0.234$}} & \multicolumn{1}{r}{} & \multicolumn{1}{r}{} \\
 & \multicolumn{1}{r}{\nopagebreak $m$}  & \multicolumn{1}{r}{$-0.869$} & \multicolumn{1}{r}{$0.466$} & \multicolumn{1}{r}{$0.605$} & \multicolumn{1}{r}{$-0.337$} & \multicolumn{1}{r}{$0.557$} & \multicolumn{1}{r}{} & \multicolumn{1}{r}{} \\
 & \multicolumn{1}{r}{\nopagebreak $\mu$}  & \multicolumn{1}{r}{$-0.612$} & \multicolumn{1}{r}{$0.188$} & \multicolumn{1}{r}{$0.519$} & \multicolumn{1}{r}{$-0.194$} & \multicolumn{1}{r}{$0.220$} & \multicolumn{1}{r}{} & \multicolumn{1}{r}{} \\
\multicolumn{1}{r}{\rule{0pt}{1.1\normalbaselineskip}\color{lang2}{Python}} & \multicolumn{1}{r}{\nopagebreak $p$}  & \multicolumn{1}{r}{\color{sstrong}{$0.000$}} & \multicolumn{1}{r}{\color{smedium}{$0.042$}} & \multicolumn{1}{r}{$0.875$} & \multicolumn{1}{r}{\color{sstrong}{$0.006$}} & \multicolumn{1}{r}{$0.334$} & \multicolumn{1}{r}{\color{sstrong}{$0.007$}} & \multicolumn{1}{r}{} \\
 & \multicolumn{1}{r}{\nopagebreak $\delta$}  & \multicolumn{1}{r}{\color{lang1}{$-0.604$}} & \multicolumn{1}{r}{\color{lang1}{$-0.265$}} & \multicolumn{1}{r}{\color{lang1}{$-0.021$}} & \multicolumn{1}{r}{\color{lang1}{$-0.423$}} & \multicolumn{1}{r}{\color{lang2}{$0.048$}} & \multicolumn{1}{r}{\color{lang1}{$-0.314$}} & \multicolumn{1}{r}{} \\
 & \multicolumn{1}{r}{\nopagebreak $m$}  & \multicolumn{1}{r}{$-0.960$} & \multicolumn{1}{r}{$-0.733$} & \multicolumn{1}{r}{$-0.016$} & \multicolumn{1}{r}{$-0.859$} & \multicolumn{1}{r}{$-0.100$} & \multicolumn{1}{r}{$-0.714$} & \multicolumn{1}{r}{} \\
 & \multicolumn{1}{r}{\nopagebreak $\mu$}  & \multicolumn{1}{r}{$-0.730$} & \multicolumn{1}{r}{$-0.378$} & \multicolumn{1}{r}{$-0.018$} & \multicolumn{1}{r}{$-0.446$} & \multicolumn{1}{r}{$0.003$} & \multicolumn{1}{r}{$-0.319$} & \multicolumn{1}{r}{} \\
\multicolumn{1}{r}{\rule{0pt}{1.1\normalbaselineskip}\color{lang2}{Ruby}} & \multicolumn{1}{r}{\nopagebreak $p$}  & \multicolumn{1}{r}{\color{sstrong}{$0.000$}} & \multicolumn{1}{r}{\color{smedium}{$0.016$}} & \multicolumn{1}{r}{$0.695$} & \multicolumn{1}{r}{\color{sstrong}{$0.003$}} & \multicolumn{1}{r}{$0.233$} & \multicolumn{1}{r}{\color{sstrong}{$0.001$}} & \multicolumn{1}{r}{$0.090$} \\
 & \multicolumn{1}{r}{\nopagebreak $\delta$}  & \multicolumn{1}{r}{\color{lang1}{$-0.609$}} & \multicolumn{1}{r}{\color{lang1}{$-0.360$}} & \multicolumn{1}{r}{\color{lang1}{$-0.042$}} & \multicolumn{1}{r}{\color{lang1}{$-0.532$}} & \multicolumn{1}{r}{\color{lang1}{$-0.059$}} & \multicolumn{1}{r}{\color{lang1}{$-0.396$}} & \multicolumn{1}{r}{\color{lang1}{$-0.100$}} \\
 & \multicolumn{1}{r}{\nopagebreak $m$}  & \multicolumn{1}{r}{$-0.973$} & \multicolumn{1}{r}{$-0.855$} & \multicolumn{1}{r}{$0.076$} & \multicolumn{1}{r}{$-0.961$} & \multicolumn{1}{r}{$-0.343$} & \multicolumn{1}{r}{$-0.831$} & \multicolumn{1}{r}{$-0.156$} \\
 & \multicolumn{1}{r}{\nopagebreak $\mu$}  & \multicolumn{1}{r}{$-0.659$} & \multicolumn{1}{r}{$-0.545$} & \multicolumn{1}{r}{$-0.030$} & \multicolumn{1}{r}{$-0.543$} & \multicolumn{1}{r}{$-0.115$} & \multicolumn{1}{r}{$-0.468$} & \multicolumn{1}{r}{$-0.173$} \\
\bottomrule 
\end{tabular}

  \caption{Frequentist reanalysis of \rosetta's Rosetta Code data about the performance of 8 programming languages.
    For each comparison of language $\ell_1$ ({\color{lang1} column header}) with language $\ell_2$ ({\color{lang2}row header}),
    the table reports:
    the $p$-value of a Wilcoxon signed-rank test comparing the running times of programs in $\ell_1$ and in $\ell_2$
    (colored differently if {\color{sstrong}$p < 0.01$} or {\color{sweak}$0.01 \leq p < 0.05$});
    Cliff's $\delta$ effect size,
    which is {\color{lang1} negative} if $\ell_1$ tends to be faster, and {\color{lang2} positive} if $\ell_2$  tends to be faster;
    and the median $m$ and mean $\mu$ of the \emph{inverse speedup} of $\ell_1$ vs.~$\ell_2$ (again, negative values indicate that $\ell_1$ is faster on average, and positive values that $\ell_2$ is).}
  \label{tab:freq_R}
\end{table*}

\begin{table*}[!h]
  \setlength{\tabcolsep}{5pt}  
  \centering
  \footnotesize
  \begin{tabular}{llccccccc}
\toprule
\textsc{language} & \textsc{correction} & \multicolumn{1}{r}{\color{lang1}{C}} & \multicolumn{1}{r}{\color{lang1}{C\#}} & \multicolumn{1}{r}{\color{lang1}{F\#}} & \multicolumn{1}{r}{\color{lang1}{Go}} & \multicolumn{1}{r}{\color{lang1}{Haskell}} & \multicolumn{1}{r}{\color{lang1}{Java}} & \multicolumn{1}{r}{\color{lang1}{Python}} \\ 
\midrule
\multicolumn{1}{r}{\nopagebreak \color{lang2}{C\#}} & \multicolumn{1}{r}{\nopagebreak --}  & \multicolumn{1}{r}{\color{sstrong}{$0.000$}} & \multicolumn{1}{r}{} & \multicolumn{1}{r}{} & \multicolumn{1}{r}{} & \multicolumn{1}{r}{} & \multicolumn{1}{r}{} & \multicolumn{1}{r}{} \\
 & \multicolumn{1}{r}{\nopagebreak B}  & \multicolumn{1}{r}{\color{sstrong}{$0.007$}} & \multicolumn{1}{r}{} & \multicolumn{1}{r}{} & \multicolumn{1}{r}{} & \multicolumn{1}{r}{} & \multicolumn{1}{r}{} & \multicolumn{1}{r}{} \\
 & \multicolumn{1}{r}{\nopagebreak H}  & \multicolumn{1}{r}{\color{sstrong}{$0.006$}} & \multicolumn{1}{r}{} & \multicolumn{1}{r}{} & \multicolumn{1}{r}{} & \multicolumn{1}{r}{} & \multicolumn{1}{r}{} & \multicolumn{1}{r}{} \\
 & \multicolumn{1}{r}{\nopagebreak B-H}  & \multicolumn{1}{r}{\color{sstrong}{$0.001$}} & \multicolumn{1}{r}{} & \multicolumn{1}{r}{} & \multicolumn{1}{r}{} & \multicolumn{1}{r}{} & \multicolumn{1}{r}{} & \multicolumn{1}{r}{} \\
\multicolumn{1}{r}{\rule{0pt}{1.1\normalbaselineskip}\color{lang2}{F\#}} & \multicolumn{1}{r}{\nopagebreak --}  & \multicolumn{1}{r}{\color{smedium}{$0.013$}} & \multicolumn{1}{r}{$0.182$} & \multicolumn{1}{r}{} & \multicolumn{1}{r}{} & \multicolumn{1}{r}{} & \multicolumn{1}{r}{} & \multicolumn{1}{r}{} \\
 & \multicolumn{1}{r}{\nopagebreak B}  & \multicolumn{1}{r}{$0.358$} & \multicolumn{1}{r}{$1.000$} & \multicolumn{1}{r}{} & \multicolumn{1}{r}{} & \multicolumn{1}{r}{} & \multicolumn{1}{r}{} & \multicolumn{1}{r}{} \\
 & \multicolumn{1}{r}{\nopagebreak H}  & \multicolumn{1}{r}{$0.205$} & \multicolumn{1}{r}{$1.000$} & \multicolumn{1}{r}{} & \multicolumn{1}{r}{} & \multicolumn{1}{r}{} & \multicolumn{1}{r}{} & \multicolumn{1}{r}{} \\
 & \multicolumn{1}{r}{\nopagebreak B-H}  & \multicolumn{1}{r}{\color{smedium}{$0.028$}} & \multicolumn{1}{r}{$0.232$} & \multicolumn{1}{r}{} & \multicolumn{1}{r}{} & \multicolumn{1}{r}{} & \multicolumn{1}{r}{} & \multicolumn{1}{r}{} \\
\multicolumn{1}{r}{\rule{0pt}{1.1\normalbaselineskip}\color{lang2}{Go}} & \multicolumn{1}{r}{\nopagebreak --}  & \multicolumn{1}{r}{\color{sstrong}{$0.000$}} & \multicolumn{1}{r}{\color{sstrong}{$0.006$}} & \multicolumn{1}{r}{\color{smedium}{$0.030$}} & \multicolumn{1}{r}{} & \multicolumn{1}{r}{} & \multicolumn{1}{r}{} & \multicolumn{1}{r}{} \\
 & \multicolumn{1}{r}{\nopagebreak B}  & \multicolumn{1}{r}{\color{sstrong}{$0.000$}} & \multicolumn{1}{r}{$0.169$} & \multicolumn{1}{r}{$0.849$} & \multicolumn{1}{r}{} & \multicolumn{1}{r}{} & \multicolumn{1}{r}{} & \multicolumn{1}{r}{} \\
 & \multicolumn{1}{r}{\nopagebreak H}  & \multicolumn{1}{r}{\color{sstrong}{$0.000$}} & \multicolumn{1}{r}{$0.121$} & \multicolumn{1}{r}{$0.394$} & \multicolumn{1}{r}{} & \multicolumn{1}{r}{} & \multicolumn{1}{r}{} & \multicolumn{1}{r}{} \\
 & \multicolumn{1}{r}{\nopagebreak B-H}  & \multicolumn{1}{r}{\color{sstrong}{$0.000$}} & \multicolumn{1}{r}{\color{smedium}{$0.018$}} & \multicolumn{1}{r}{$0.053$} & \multicolumn{1}{r}{} & \multicolumn{1}{r}{} & \multicolumn{1}{r}{} & \multicolumn{1}{r}{} \\
\multicolumn{1}{r}{\rule{0pt}{1.1\normalbaselineskip}\color{lang2}{Haskell}} & \multicolumn{1}{r}{\nopagebreak --}  & \multicolumn{1}{r}{\color{sstrong}{$0.000$}} & \multicolumn{1}{r}{$0.142$} & \multicolumn{1}{r}{$0.929$} & \multicolumn{1}{r}{\color{smedium}{$0.025$}} & \multicolumn{1}{r}{} & \multicolumn{1}{r}{} & \multicolumn{1}{r}{} \\
 & \multicolumn{1}{r}{\nopagebreak B}  & \multicolumn{1}{r}{\color{sstrong}{$0.001$}} & \multicolumn{1}{r}{$1.000$} & \multicolumn{1}{r}{$1.000$} & \multicolumn{1}{r}{$0.706$} & \multicolumn{1}{r}{} & \multicolumn{1}{r}{} & \multicolumn{1}{r}{} \\
 & \multicolumn{1}{r}{\nopagebreak H}  & \multicolumn{1}{r}{\color{sstrong}{$0.001$}} & \multicolumn{1}{r}{$1.000$} & \multicolumn{1}{r}{$1.000$} & \multicolumn{1}{r}{$0.353$} & \multicolumn{1}{r}{} & \multicolumn{1}{r}{} & \multicolumn{1}{r}{} \\
 & \multicolumn{1}{r}{\nopagebreak B-H}  & \multicolumn{1}{r}{\color{sstrong}{$0.000$}} & \multicolumn{1}{r}{$0.189$} & \multicolumn{1}{r}{$0.929$} & \multicolumn{1}{r}{\color{smedium}{$0.047$}} & \multicolumn{1}{r}{} & \multicolumn{1}{r}{} & \multicolumn{1}{r}{} \\
\multicolumn{1}{r}{\rule{0pt}{1.1\normalbaselineskip}\color{lang2}{Java}} & \multicolumn{1}{r}{\nopagebreak --}  & \multicolumn{1}{r}{\color{sstrong}{$0.000$}} & \multicolumn{1}{r}{$0.140$} & \multicolumn{1}{r}{\color{sstrong}{$0.008$}} & \multicolumn{1}{r}{$0.451$} & \multicolumn{1}{r}{$0.064$} & \multicolumn{1}{r}{} & \multicolumn{1}{r}{} \\
 & \multicolumn{1}{r}{\nopagebreak B}  & \multicolumn{1}{r}{\color{sstrong}{$0.002$}} & \multicolumn{1}{r}{$1.000$} & \multicolumn{1}{r}{$0.214$} & \multicolumn{1}{r}{$1.000$} & \multicolumn{1}{r}{$1.000$} & \multicolumn{1}{r}{} & \multicolumn{1}{r}{} \\
 & \multicolumn{1}{r}{\nopagebreak H}  & \multicolumn{1}{r}{\color{sstrong}{$0.002$}} & \multicolumn{1}{r}{$1.000$} & \multicolumn{1}{r}{$0.130$} & \multicolumn{1}{r}{$1.000$} & \multicolumn{1}{r}{$0.709$} & \multicolumn{1}{r}{} & \multicolumn{1}{r}{} \\
 & \multicolumn{1}{r}{\nopagebreak B-H}  & \multicolumn{1}{r}{\color{sstrong}{$0.000$}} & \multicolumn{1}{r}{$0.189$} & \multicolumn{1}{r}{\color{smedium}{$0.018$}} & \multicolumn{1}{r}{$0.505$} & \multicolumn{1}{r}{$0.100$} & \multicolumn{1}{r}{} & \multicolumn{1}{r}{} \\
\multicolumn{1}{r}{\rule{0pt}{1.1\normalbaselineskip}\color{lang2}{Python}} & \multicolumn{1}{r}{\nopagebreak --}  & \multicolumn{1}{r}{\color{sstrong}{$0.000$}} & \multicolumn{1}{r}{\color{smedium}{$0.042$}} & \multicolumn{1}{r}{$0.875$} & \multicolumn{1}{r}{\color{sstrong}{$0.006$}} & \multicolumn{1}{r}{$0.334$} & \multicolumn{1}{r}{\color{sstrong}{$0.007$}} & \multicolumn{1}{r}{} \\
 & \multicolumn{1}{r}{\nopagebreak B}  & \multicolumn{1}{r}{\color{sstrong}{$0.000$}} & \multicolumn{1}{r}{$1.000$} & \multicolumn{1}{r}{$1.000$} & \multicolumn{1}{r}{$0.180$} & \multicolumn{1}{r}{$1.000$} & \multicolumn{1}{r}{$0.198$} & \multicolumn{1}{r}{} \\
 & \multicolumn{1}{r}{\nopagebreak H}  & \multicolumn{1}{r}{\color{sstrong}{$0.000$}} & \multicolumn{1}{r}{$0.502$} & \multicolumn{1}{r}{$1.000$} & \multicolumn{1}{r}{$0.122$} & \multicolumn{1}{r}{$1.000$} & \multicolumn{1}{r}{$0.128$} & \multicolumn{1}{r}{} \\
 & \multicolumn{1}{r}{\nopagebreak B-H}  & \multicolumn{1}{r}{\color{sstrong}{$0.000$}} & \multicolumn{1}{r}{$0.069$} & \multicolumn{1}{r}{$0.908$} & \multicolumn{1}{r}{\color{smedium}{$0.018$}} & \multicolumn{1}{r}{$0.389$} & \multicolumn{1}{r}{\color{smedium}{$0.018$}} & \multicolumn{1}{r}{} \\
\multicolumn{1}{r}{\rule{0pt}{1.1\normalbaselineskip}\color{lang2}{Ruby}} & \multicolumn{1}{r}{\nopagebreak --}  & \multicolumn{1}{r}{\color{sstrong}{$0.000$}} & \multicolumn{1}{r}{\color{smedium}{$0.016$}} & \multicolumn{1}{r}{$0.695$} & \multicolumn{1}{r}{\color{sstrong}{$0.003$}} & \multicolumn{1}{r}{$0.233$} & \multicolumn{1}{r}{\color{sstrong}{$0.001$}} & \multicolumn{1}{r}{$0.090$} \\
 & \multicolumn{1}{r}{\nopagebreak B}  & \multicolumn{1}{r}{\color{sstrong}{$0.001$}} & \multicolumn{1}{r}{$0.441$} & \multicolumn{1}{r}{$1.000$} & \multicolumn{1}{r}{$0.083$} & \multicolumn{1}{r}{$1.000$} & \multicolumn{1}{r}{\color{smedium}{$0.025$}} & \multicolumn{1}{r}{$1.000$} \\
 & \multicolumn{1}{r}{\nopagebreak H}  & \multicolumn{1}{r}{\color{sstrong}{$0.001$}} & \multicolumn{1}{r}{$0.236$} & \multicolumn{1}{r}{$1.000$} & \multicolumn{1}{r}{$0.062$} & \multicolumn{1}{r}{$1.000$} & \multicolumn{1}{r}{\color{smedium}{$0.020$}} & \multicolumn{1}{r}{$0.900$} \\
 & \multicolumn{1}{r}{\nopagebreak B-H}  & \multicolumn{1}{r}{\color{sstrong}{$0.000$}} & \multicolumn{1}{r}{\color{smedium}{$0.031$}} & \multicolumn{1}{r}{$0.748$} & \multicolumn{1}{r}{\color{smedium}{$0.010$}} & \multicolumn{1}{r}{$0.283$} & \multicolumn{1}{r}{\color{sstrong}{$0.004$}} & \multicolumn{1}{r}{$0.133$} \\
\bottomrule 
\end{tabular}

  \caption{Frequentist reanalysis of \rosetta's Rosetta Code data about the performance of 8 programming languages.
    For each comparison of language $\ell_1$ ({\color{lang1} column header}) with language $\ell_2$ ({\color{lang2}row header}),
    the table reports the $p$-value of a Wilcoxon signed-rank test---comparing the running times of programs in $\ell_1$ and in $\ell_2$---corrected
    with different methods:
    no correction (--, identical to the $p$-values in \autoref{tab:freq_R}),
    Bonferroni correction (B),
    Holm correction (H),
    Benjamini-Hochberg correction (B-H).
    Values are colored if {\color{sstrong}$p < 0.01$} or {\color{sweak}$0.01 \leq p < 0.05$}.}
  \label{tab:freq_corrected_R}
\end{table*}

\myparagraph{Frequentist analysis results.}
\autoref{tab:freq_R} shows the results of \rosetta's frequentist analysis using colors to mark $p$-values that are {\color{sstrong} statistically significant at level $0.01$} ($p < 0.01$) and {\color{sweak} statistically significant at level $0.05$} ($0.01 \leq p < 0.05$); 
for significant comparisons, the effects $\delta$ are colored according to which language (the one in the {\color{lang1} column} header, or in the {\color{lang2} row} header) is faster, that is whether $\delta$ is negative or positive.
\autoref{tab:freq_R} also reports summary statistics of the same data: the median $m$ and mean $\mu$ inverse speedup $\widehat{S}_{1,2}(t)$ across all tasks $t \in T_{1,2}$. 

Take for example the comparison between {\color{lang1} F\#} and {\color{lang2} Go} (column \#5, row \#3). The $p$-value is colored in dark blue because $0.01 < {\color{sweak} p = 0.025} < 0.05$, and hence the performance difference between the two languages is statistically significant at level $0.05$.
The effect size $\delta$ is colored in green because it is positive ${\color{lang2} \delta = 0.408} > 0$ meaning that the language on the \emph{row header} ({\color{lang2} Go}, also in green) was faster on average.

\begin{figure*}[!tb]
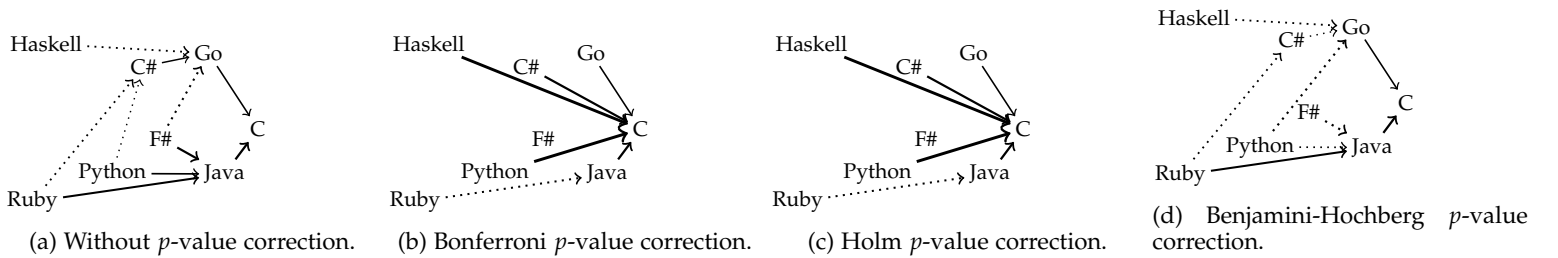

  \centering
\begin{adjustwidth}{-13mm}{-13mm}
  \begin{subfigure}[b]{0.24\linewidth}
    \rosettagraph{graph_freq_value_R_blayout}
    \caption{Without $p$-value correction.}
    \label{fig:graph_freq_R_nocorrection}
  \end{subfigure}
  \begin{subfigure}[b]{0.24\linewidth}
    \rosettagraph{graph_freq_bonferroni_R_blayout}
    \caption{Bonferroni $p$-value correction.}
    \label{fig:graph_freq_R_bonferroni}
  \end{subfigure}
  \begin{subfigure}[b]{0.24\linewidth}
    \rosettagraph{graph_freq_holm_R_blayout}
    \caption{Holm $p$-value correction.}
    \label{fig:graph_freq_R_holm}
  \end{subfigure}
  \begin{subfigure}[b]{0.24\linewidth}
    \rosettagraph{graph_freq_BH_R_blayout}
    \caption{Benjamini-Hochberg $p$-value correction.}
    \label{fig:graph_freq_R_BH}
  \end{subfigure}
\end{adjustwidth}
\caption{Language relationship graphs summarizing the frequentist reanalysis of \rosetta, with different $p$-value correction methods.
  In every graph, an arrow from node $\ell_1$ to node $\ell_2$ indicates that the $p$-value comparing performance data between the two languages is $p < 0.01$ (solid line) or $0.01 \leq p < 0.05$ (dotted line), and Cliff's $\delta$ effect size indicates that $\ell_2$ tends to be faster;
  the thickness of the arrows is proportional to the absolute value of Cliff's $\delta$.}
  \label{fig:graph_freq_R}
\end{figure*}

The \emph{language relationship graph} in \autoref{fig:graph_freq_R} summarizes all pairwise comparisons: nodes are languages; an arrow from $\ell_1$ to $\ell_2$ denotes that the difference between the two languages is significant ($p < 0.05$), and goes from the slower to the faster language according to the sign of $\delta$; 
dotted arrows denote significant differences at level $0.05$ but not at level $0.01$;
the thickness of the arrows is proportional to the absolute value of the effect $\delta$.
To make the graph less cluttered, after checking that the induced relation is transitive, we remove the arrows that are subsumed by transitivity.

\subsubsection{Multiple Comparisons}
\label{sec:rosetta-multiple}

Conspicuously absent from \rosetta's analysis is how to deal with the \emph{multiple comparisons} problem~\cite{Miller-multistats}
(also known as \emph{multiple tests} problem).
As the name suggests, the problem occurs when several statistical tests are applied to the same dataset.
If each test rejects the null hypothesis whenever $p < \alpha$,
the probability of committing a type~1 error (see \autoref{sec:type12-sm-errors})
in \emph{at least one} test can grow as high as $1 - (1 - \alpha)^N$
(where $N$ is the number of independent tests),
which grows to 1 as $N$ increases.

In a frequentist settings there are two main approaches to address the multiple comparisons problem~\cite{hitchhiker-journal,Miller-multistats}:
\begin{itemize}
\item adjust $p$-values to compensate for the increased probability of error; or 
\item use a so-called ``omnibus'' test, which can deal with multiple comparisons at the same time.
\end{itemize}
In this section we explore both alternatives on the Rosetta Code data.

\myparagraph{P-value adjustment.}
The intuition behind $p$-value adjustment (also: correction)
is to reduce the $p$-value threshold to compensate for the increased probability of ``spurious''
null hypothesis rejections which are simply a result of multiple tests.
For example, the Bonferroni adjustment works as follows: if we are performing $N$ tests on the same data
and aim for a $\alpha$ significance level,
we will reject the null hypothesis in a test yielding $p$-value $p$ only if $p \cdot N < \alpha$---thus effectively tightening the significance level.

There is no uniform view about when and how to apply $p$-value adjustments~\cite{bonfvsh,hitchhiker-journal,Bonferroni3,Bonferroni1,Bonferroni2,mcorr-howto};
and there are different procedures for performing such adjustments, which differ in how strictly they correct the $p$-values
(with Bonferroni's being the strictest adjustment as it assumes a ``worst-case'' scenario).
To provide as broad a view as possible, we applied three widely used $p$-value corrections,
in decreasing strictness:
  Bonferroni's~\cite{bonfvsh},
  Holm's~\cite{bonfvsh}, and 
  Benjamini-Hochberg's~\cite{BH-original}.

\autoref{tab:freq_corrected_R} shows the results compared to the unadjusted $p$-values;
and \autoref{fig:graph_freq_R} shows how the language relationship graph changes with different significance levels.
Even the mildest correction reduces the number of significant language performance differences that we can claim,
and strict corrections drastically reduce it:
the 17 language pairs such that $p < 0.05$ without adjustment
become 15 pairs using Benjamini-Hochberg's,
and only 7 pairs using Holm's or Bonferroni's.

\myparagraph{Multiple comparison testing.}
Frequentist  statistics also offers tests to simultaneously compare samples from several groups in a way that gets around the multiple comparisons problem.
The Kruskal-Wallis test~\cite{HW} is one such test
that is applicable to nonparametric data, since it is an extension of the Mann-Whitney test to more than two groups, or a nonparametric version of ANOVA on ranks.

To apply Kruskal-Wallis, the samples from the various groups must be directly comparable.
The frequentist analysis done so far has focused on \emph{pairwise comparisons} resulting in inverse speedup ratios;
in contrast, we have to consider absolute running times to be able to run the Kruskal-Wallis test.
Unfortunately, this means that only the Rosetta Code tasks for which \emph{all 8 languages} have an implementation can be included in the analysis (using the notation of \autoref{sec:data-testing}: $\bigcap_{\ell} T(\ell)$);
otherwise, we would be including running times on unrelated tasks, which are not comparable in a meaningful way.
This restriction leaves us with a mere 6 data points per language---for a total of $24 = 6 \cdot 8$ running times.

On this dataset, the Kruskal-Wallis test indicates that there are statistically significant differences ($p = 0.002 < 0.05$);
but the test cannot specify \emph{which} language pairs differ significantly.
To find these out, it is customary to run a post-hoc analysis.
We tried several that are appropriate for this setting:
\begin{description}
\item[MC:] a multiple-comparison between treatments~\cite[Sec.~8.3.3]{SiegelCastellan}
\item[Nemenyi:] Nemenyi's test of multiple comparisons~\cite{Zar}
\item[Dunn:] Dunn's test of multiple comparisons~\cite{Zar} with $p$-values adjusted using Benjamini-Hochberg
\item[PW:] a pairwise version of the Wilcoxon test with $p$-values adjusted using Benjamini-Hochberg~\cite{HW}
\end{description}

The number of significantly different language pairs are as follows:
\begin{center}
\begin{tabular}{crr}
  \toprule
  \textsc{post hoc} & 95\% level & 99\% level  \\
  \midrule
  MC & 2 & 0 \\
  Nemenyi & 2 & 0 \\
  Dunn  & 3 & 0 \\
  PW & 5 & 0 \\
  \bottomrule
\end{tabular}
\end{center}
A closer look into the results shows that 
all the significant comparisons involve C and another language, with the exception of the comparisons between Go and Python, and Go and Ruby,
which the pariwise Wilcoxon classifies as significant at the 95\% level;
no comparison is significant at the 99\% level.

\subsubsection{Summary and Criticism of Frequentist Statistics}
Analyzing significance using hypothesis testing is problematic for a number of general reasons.
First, a null hypothesis is an oversimplification since it forces a binary choice between the extreme case of \emph{no effects}---which is unlikely to happen in practice---versus \emph{some effects}.
Second, $p$-values are probabilities conditional on the null hypothesis being true;
using them to infer whether the null hypothesis itself is likely true is unsound in general.

Even if we brush these general methodological problems aside,
null-hy\-poth\-e\-sis testing becomes even murkier when dealing with multiple tests.
If we choose to ignore the multiple tests problem, we may end up with incorrect results that overestimate the weight of evidence;
if we decide to add corrections it is debatable which one is the most appropriate,
with a conflicting choice between ``strict'' corrections---safer from a methodological viewpoint, but also likely to strike down any significant findings---and ``weak'' corrections---salvaging significance to a larger degree, but possibly arbitrarily chosen.

Using effect sizes is a definite improvement over $p$-values:
effect sizes are continuous measures rather than binary alternatives;
and are unconditional summary statistics of the experimental data.
Nonetheless, they remain elementary statistics that provide limited information.
In our case study, especially in the cases where the results of a language comparison are somewhat borderline or inconclusive,
it is unclear whether an effect size provides a more reliable measure than statistics
such as mean or median (even when one takes into consideration common language effect size statistics~\cite{hitchhiker-icse11}, normally referred to as Vargha-Delaney effect size~\cite{VD-paper}). Besides, merely summarizing the experimental data is a very primitive way of modeling it,
as it cannot easily accommodate different assumptions to study how they would change our quantitative interpretation of the data.
Finally, frequentist best practices normally recommend to use effect size \emph{in addition to} null-hypothesis testing;
which brings additional problems whenever a test does not reach significance, but the corresponding effect size is non-negligible.

As some examples that expose these general weaknesses,
consider the four language comparisons C\# vs.\ F\#,  F\# vs.\ Python, F\# vs.\ Ruby, and Haskell vs.\ Python.
According to the $p$-values the four comparisons are inconclusive. The effect sizes tell a different story:
they suggest a sizable speed advantage of C\# over F\# ($\delta = -0.223$),
and a small to negligible difference in favor of F\# over Python ($\delta = -0.021$), of F\# over Ruby ($\delta = -0.042$), and of Python over Haskell ($\delta = 0.048$).
Other summary statistics make the picture even more confusing:
while median and mean agree with the effect size in the F\# vs.~C\# and F\# vs.~Python comparisons,
the median suggests that Ruby is slightly faster than F\# ($m = 0.076$), and that Haskell is faster than Python ($m = -0.1$);
whereas mean and effect size suggest that F\# ($\mu = -0.03$) and Python ($\mu = 0.003$) are slightly faster.
We might argue about the merits of one statistic over the other,
but there is no clear-cut way to fine-tune the analysis without arbitrarily complicating it.
In contrast, with Bayesian statistics we can obtain nuanced, yet natural-to-interpret,
analyses of the very same experimental data---as we demonstrate next.

\subsection{Bayesian Reanalysis}
\label{sec:bayesian-rosetta}

We analyze \rosetta's performance data using an explicit Bayesian model.
For each language pair $\ell_1, \ell_2$, the model expresses the probability
that one language achieves a certain speedup $s$ over the other language.
Speedup $s$ is measured by the inverse speedup ratio \eqref{eq:ratio-def}
ranging over the interval $(-1, +1)$.

\subsubsection{Statistical Model}
\label{sec:rosetta:bayes-model}
To apply Bayes' theorem, we need a likelihood---linking
observed data to inferred data---and a prior---expressing our initial assumption
about the speedup.

\myparagraph{Likelihood.}
The likelihood $\like[d, s]$ weighs how likely we are to observe an inverse speedup $d$ assuming that the ``real'' inverse speedup is $s$.
Using the terminology of \autoref{sec:bayes-overview}, $d$ is the \emph{data} and \emph{s} is the \emph{hypothesis} of probabilistic inference.
In this case, the likelihood essentially models possible imprecisions in the experimental data;
thus, we express it in terms of the difference $d - s$ between observed and the ``real'' speedup:
\[
  d - s \sim \dist{N}(0, \sigma)\,,
\]
that is the difference is normally distributed with mean zero (no systematic error) and unknown standard deviation.
This just models experimental noise that might affect the measured speedups (and corresponds to all sorts of approximations and effects that we cannot control for) so that they do not coincide with the ``real'' speedups.

\myparagraph{Priors.}
The prior $\prior[s]$ models initial assumptions on what a plausible speedup for the pair of languages might be.
We consider three different priors, from uniform to biased:
\begin{enumerate}
\item a \emph{uniform} prior $\up(-1, 1)$---a uniform distribution over $(-1, 1)$;
\item a weak unbiased \emph{centered normal} prior $\np(0, \sigma_1)$---a normal distribution with mean $0$ and standard deviation $\sigma_1$, truncated to have support $(-1, 1)$;
\item a biased \emph{shifted normal} prior $\np(\mu_2, \sigma_2)$---a normal distribution with mean $\mu_2$ and standard deviation $\sigma_2$, truncated to have support $(-1, 1)$;
\end{enumerate}
Picking different priors corresponds to
assessing the impact of different assumptions on the data under analysis.
Unlike the uniform prior, the normal priors have to be based on some data other than \rosetta's.
To this end we use the Computer Language Benchmarks Game~\cite{benchmarks} (for brevity, \emph{\bench}).
The data from \bench are comparable to those from \rosetta as they also consist of curated selections of collectively written solutions to well-defined programming tasks running on the same input and refined over a significant stretch of time; however, \bench was developed independently of Rosetta Code, which makes it a complementary source of data.

In the centered normal prior, we set $\sigma_1$ to the largest absolute inverse speedup value observed in \bench between $\ell_1$ and $\ell_2$.
This represents the assumption that the largest speedups observed in \bench give a range of plausible maximum values for speedups observed in Rosetta Code.
This is still a very weak prior since maxima will mostly be much larger than typical values;
and there is no bias because the distribution mean is zero.

In the shifted normal prior, we set $\mu_2$ and $\sigma_2$ to the mean and standard deviation of \bench's optimal data.
The resulting distribution is biased towards \bench's data, as it uses \bench's performance data to shape the prior assumptions;
if $\mu_2 \gg 0$ or $\mu_2 \ll 0$, the priors are biased in favor of one of the two languages being faster.
In most cases, however, the biased priors still allow for a wide range of speedups; and, in any case, they can still be swamped by strong evidence in \rosetta's data.

\myparagraph{Posteriors and uncertainty intervals.}
For each pair of languages $\ell_1$ and $\ell_2$, for each prior,
applying Bayes' theorem gives a posterior distribution $\post_{1,2}(s)$, which assigns a probability to any possible value $s$ of inverse speedup.
We summarize the posterior using descriptive statistics:
\begin{enumerate}
\item 95\% and 99\% uncertainty intervals;

\item median $m$ and mean $\mu$ of the posterior.
\end{enumerate}

Since we are trying to establish which language, $\ell_1$ or $\ell_2$, is faster according to our experiments, it is useful to check whether the uncertainty intervals include the origin or not.
For 95\% probability, consider the leftmost uncertainty interval $(c_{0\%}, c_{95\%})$ and the rightmost uncertainty interval $(c_{5\%}, c_{100\%})$; that is, $c_x$ is the value such that the posterior probability of drawing an inverse speedup less than $c_x$ is $x$: $\int_{-1}^{c_x} \post_{1,2}(s) \de s < x$.
If $c_{95\%} < 0$ there is a 95\% chance that $\ell_1$ is at least $c_{95\%}$  faster than $\ell_2$ according to the posterior; conversely, if $0 < c_{5\%}$ there is a 95\% chance that $\ell_2$ is at least  $c_{5\%}$ faster;\footnote{Since $c_{5\%} \leq c_{95\%}$, the two conditions are mutually exclusive.}
if neither is the case, the comparison between the two languages is inconclusive at a 95\% certainty level.
A similar reading applies to the 99\% uncertainty intervals.

\begin{sidewaystable*}
\centering
\scriptsize
\setlength{\tabcolsep}{2pt}
\begin{tabular}{llccccccccccccccccccccc}
\toprule
& & \multicolumn{3}{c}{\color{lang1}{C}} & \multicolumn{3}{c}{\color{lang1}{C\#}} & \multicolumn{3}{c}{\color{lang1}{F\#}} & \multicolumn{3}{c}{\color{lang1}{Go}} & \multicolumn{3}{c}{\color{lang1}{Haskell}} & \multicolumn{3}{c}{\color{lang1}{Java}} & \multicolumn{3}{c}{\color{lang1}{Python}} \\ \cmidrule(lr){3-5}\cmidrule(lr){6-8}\cmidrule(lr){9-11}\cmidrule(lr){12-14}\cmidrule(lr){15-17}\cmidrule(lr){18-20}\cmidrule(lr){21-23}
\textsc{language} & \textsc{measure} & \multicolumn{1}{c}{$\mathcal{U}$} & \multicolumn{1}{c}{$\mathcal{N}$} & \multicolumn{1}{c}{$\mathcal{S}$} & \multicolumn{1}{c}{$\mathcal{U}$} & \multicolumn{1}{c}{$\mathcal{N}$} & \multicolumn{1}{c}{$\mathcal{S}$} & \multicolumn{1}{c}{$\mathcal{U}$} & \multicolumn{1}{c}{$\mathcal{N}$} & \multicolumn{1}{c}{$\mathcal{S}$} & \multicolumn{1}{c}{$\mathcal{U}$} & \multicolumn{1}{c}{$\mathcal{N}$} & \multicolumn{1}{c}{$\mathcal{S}$} & \multicolumn{1}{c}{$\mathcal{U}$} & \multicolumn{1}{c}{$\mathcal{N}$} & \multicolumn{1}{c}{$\mathcal{S}$} & \multicolumn{1}{c}{$\mathcal{U}$} & \multicolumn{1}{c}{$\mathcal{N}$} & \multicolumn{1}{c}{$\mathcal{S}$} & \multicolumn{1}{c}{$\mathcal{U}$} & \multicolumn{1}{c}{$\mathcal{N}$} & \multicolumn{1}{c}{$\mathcal{S}$} \\ 
\midrule
\multicolumn{1}{r}{\nopagebreak \color{lang2}{C\#}} & \multicolumn{1}{r}{\nopagebreak \color{smedium}{95{\tiny \%}}}  & \multicolumn{1}{r}{\color{lang1}{$-0.398$}} & \multicolumn{1}{r}{\color{lang1}{$-0.392$}} & \multicolumn{1}{r}{\color{lang1}{$-0.631$}} & \multicolumn{1}{r}{} & \multicolumn{1}{r}{} & \multicolumn{1}{r}{} & \multicolumn{1}{r}{} & \multicolumn{1}{r}{} & \multicolumn{1}{r}{} & \multicolumn{1}{r}{} & \multicolumn{1}{r}{} & \multicolumn{1}{r}{} & \multicolumn{1}{r}{} & \multicolumn{1}{r}{} & \multicolumn{1}{r}{} & \multicolumn{1}{r}{} & \multicolumn{1}{r}{} & \multicolumn{1}{r}{} & \multicolumn{1}{r}{} & \multicolumn{1}{r}{} & \multicolumn{1}{r}{} \\
 & \multicolumn{1}{r}{\nopagebreak \color{sstrong}{99{\tiny \%}}}  & \multicolumn{1}{r}{\color{lang1}{$-0.307$}} & \multicolumn{1}{r}{\color{lang1}{$-0.306$}} & \multicolumn{1}{r}{\color{lang1}{$-0.610$}} & \multicolumn{1}{r}{} & \multicolumn{1}{r}{} & \multicolumn{1}{r}{} & \multicolumn{1}{r}{} & \multicolumn{1}{r}{} & \multicolumn{1}{r}{} & \multicolumn{1}{r}{} & \multicolumn{1}{r}{} & \multicolumn{1}{r}{} & \multicolumn{1}{r}{} & \multicolumn{1}{r}{} & \multicolumn{1}{r}{} & \multicolumn{1}{r}{} & \multicolumn{1}{r}{} & \multicolumn{1}{r}{} & \multicolumn{1}{r}{} & \multicolumn{1}{r}{} & \multicolumn{1}{r}{} \\
 & \multicolumn{1}{r}{\nopagebreak $m$}  & \multicolumn{1}{r}{$-0.601$} & \multicolumn{1}{r}{$-0.594$} & \multicolumn{1}{r}{$-0.678$} & \multicolumn{1}{r}{} & \multicolumn{1}{r}{} & \multicolumn{1}{r}{} & \multicolumn{1}{r}{} & \multicolumn{1}{r}{} & \multicolumn{1}{r}{} & \multicolumn{1}{r}{} & \multicolumn{1}{r}{} & \multicolumn{1}{r}{} & \multicolumn{1}{r}{} & \multicolumn{1}{r}{} & \multicolumn{1}{r}{} & \multicolumn{1}{r}{} & \multicolumn{1}{r}{} & \multicolumn{1}{r}{} & \multicolumn{1}{r}{} & \multicolumn{1}{r}{} & \multicolumn{1}{r}{} \\
 & \multicolumn{1}{r}{\nopagebreak $\mu$}  & \multicolumn{1}{r}{$-0.598$} & \multicolumn{1}{r}{$-0.592$} & \multicolumn{1}{r}{$-0.677$} & \multicolumn{1}{r}{} & \multicolumn{1}{r}{} & \multicolumn{1}{r}{} & \multicolumn{1}{r}{} & \multicolumn{1}{r}{} & \multicolumn{1}{r}{} & \multicolumn{1}{r}{} & \multicolumn{1}{r}{} & \multicolumn{1}{r}{} & \multicolumn{1}{r}{} & \multicolumn{1}{r}{} & \multicolumn{1}{r}{} & \multicolumn{1}{r}{} & \multicolumn{1}{r}{} & \multicolumn{1}{r}{} & \multicolumn{1}{r}{} & \multicolumn{1}{r}{} & \multicolumn{1}{r}{} \\
\multicolumn{1}{r}{\rule{0pt}{1.1\normalbaselineskip}\color{lang2}{F\#}} & \multicolumn{1}{r}{\nopagebreak \color{smedium}{95{\tiny \%}}}  & \multicolumn{1}{r}{\color{lang1}{$-0.355$}} & \multicolumn{1}{r}{\color{lang1}{$-0.347$}} & \multicolumn{1}{r}{\color{lang1}{$-0.411$}} & \multicolumn{1}{r}{\color{lang1}{$-0.082$}} & \multicolumn{1}{r}{\color{lang1}{$-0.073$}} & \multicolumn{1}{r}{\color{lang1}{$-0.080$}} & \multicolumn{1}{r}{} & \multicolumn{1}{r}{} & \multicolumn{1}{r}{} & \multicolumn{1}{r}{} & \multicolumn{1}{r}{} & \multicolumn{1}{r}{} & \multicolumn{1}{r}{} & \multicolumn{1}{r}{} & \multicolumn{1}{r}{} & \multicolumn{1}{r}{} & \multicolumn{1}{r}{} & \multicolumn{1}{r}{} & \multicolumn{1}{r}{} & \multicolumn{1}{r}{} & \multicolumn{1}{r}{} \\
 & \multicolumn{1}{r}{\nopagebreak \color{sstrong}{99{\tiny \%}}}  & \multicolumn{1}{r}{\color{lang1}{$-0.178$}} & \multicolumn{1}{r}{\color{lang1}{$-0.172$}} & \multicolumn{1}{r}{\color{lang1}{$-0.301$}} & \multicolumn{1}{r}{$+0.000$} & \multicolumn{1}{r}{$+0.000$} & \multicolumn{1}{r}{$+0.000$} & \multicolumn{1}{r}{} & \multicolumn{1}{r}{} & \multicolumn{1}{r}{} & \multicolumn{1}{r}{} & \multicolumn{1}{r}{} & \multicolumn{1}{r}{} & \multicolumn{1}{r}{} & \multicolumn{1}{r}{} & \multicolumn{1}{r}{} & \multicolumn{1}{r}{} & \multicolumn{1}{r}{} & \multicolumn{1}{r}{} & \multicolumn{1}{r}{} & \multicolumn{1}{r}{} & \multicolumn{1}{r}{} \\
 & \multicolumn{1}{r}{\nopagebreak $m$}  & \multicolumn{1}{r}{$-0.682$} & \multicolumn{1}{r}{$-0.661$} & \multicolumn{1}{r}{$-0.651$} & \multicolumn{1}{r}{$-0.367$} & \multicolumn{1}{r}{$-0.360$} & \multicolumn{1}{r}{$-0.311$} & \multicolumn{1}{r}{} & \multicolumn{1}{r}{} & \multicolumn{1}{r}{} & \multicolumn{1}{r}{} & \multicolumn{1}{r}{} & \multicolumn{1}{r}{} & \multicolumn{1}{r}{} & \multicolumn{1}{r}{} & \multicolumn{1}{r}{} & \multicolumn{1}{r}{} & \multicolumn{1}{r}{} & \multicolumn{1}{r}{} & \multicolumn{1}{r}{} & \multicolumn{1}{r}{} & \multicolumn{1}{r}{} \\
 & \multicolumn{1}{r}{\nopagebreak $\mu$}  & \multicolumn{1}{r}{$-0.667$} & \multicolumn{1}{r}{$-0.649$} & \multicolumn{1}{r}{$-0.647$} & \multicolumn{1}{r}{$-0.367$} & \multicolumn{1}{r}{$-0.358$} & \multicolumn{1}{r}{$-0.308$} & \multicolumn{1}{r}{} & \multicolumn{1}{r}{} & \multicolumn{1}{r}{} & \multicolumn{1}{r}{} & \multicolumn{1}{r}{} & \multicolumn{1}{r}{} & \multicolumn{1}{r}{} & \multicolumn{1}{r}{} & \multicolumn{1}{r}{} & \multicolumn{1}{r}{} & \multicolumn{1}{r}{} & \multicolumn{1}{r}{} & \multicolumn{1}{r}{} & \multicolumn{1}{r}{} & \multicolumn{1}{r}{} \\
\multicolumn{1}{r}{\rule{0pt}{1.1\normalbaselineskip}\color{lang2}{Go}} & \multicolumn{1}{r}{\nopagebreak \color{smedium}{95{\tiny \%}}}  & \multicolumn{1}{r}{\color{lang1}{$-0.358$}} & \multicolumn{1}{r}{\color{lang1}{$-0.355$}} & \multicolumn{1}{r}{\color{lang1}{$-0.363$}} & \multicolumn{1}{r}{\color{lang2}{$0.099$}} & \multicolumn{1}{r}{\color{lang2}{$0.091$}} & \multicolumn{1}{r}{\color{lang2}{$0.090$}} & \multicolumn{1}{r}{\color{lang2}{$0.187$}} & \multicolumn{1}{r}{\color{lang2}{$0.169$}} & \multicolumn{1}{r}{\color{lang2}{$0.189$}} & \multicolumn{1}{r}{} & \multicolumn{1}{r}{} & \multicolumn{1}{r}{} & \multicolumn{1}{r}{} & \multicolumn{1}{r}{} & \multicolumn{1}{r}{} & \multicolumn{1}{r}{} & \multicolumn{1}{r}{} & \multicolumn{1}{r}{} & \multicolumn{1}{r}{} & \multicolumn{1}{r}{} & \multicolumn{1}{r}{} \\
 & \multicolumn{1}{r}{\nopagebreak \color{sstrong}{99{\tiny \%}}}  & \multicolumn{1}{r}{\color{lang1}{$-0.304$}} & \multicolumn{1}{r}{\color{lang1}{$-0.300$}} & \multicolumn{1}{r}{\color{lang1}{$-0.316$}} & \multicolumn{1}{r}{\color{lang2}{$+0.001$}} & \multicolumn{1}{r}{$+0.000$} & \multicolumn{1}{r}{\color{lang2}{$+0.009$}} & \multicolumn{1}{r}{\color{lang2}{$+0.053$}} & \multicolumn{1}{r}{\color{lang2}{$+0.035$}} & \multicolumn{1}{r}{\color{lang2}{$+0.090$}} & \multicolumn{1}{r}{} & \multicolumn{1}{r}{} & \multicolumn{1}{r}{} & \multicolumn{1}{r}{} & \multicolumn{1}{r}{} & \multicolumn{1}{r}{} & \multicolumn{1}{r}{} & \multicolumn{1}{r}{} & \multicolumn{1}{r}{} & \multicolumn{1}{r}{} & \multicolumn{1}{r}{} & \multicolumn{1}{r}{} \\
 & \multicolumn{1}{r}{\nopagebreak $m$}  & \multicolumn{1}{r}{$-0.473$} & \multicolumn{1}{r}{$-0.473$} & \multicolumn{1}{r}{$-0.470$} & \multicolumn{1}{r}{$+0.386$} & \multicolumn{1}{r}{$+0.375$} & \multicolumn{1}{r}{$+0.321$} & \multicolumn{1}{r}{$+0.535$} & \multicolumn{1}{r}{$+0.513$} & \multicolumn{1}{r}{$+0.458$} & \multicolumn{1}{r}{} & \multicolumn{1}{r}{} & \multicolumn{1}{r}{} & \multicolumn{1}{r}{} & \multicolumn{1}{r}{} & \multicolumn{1}{r}{} & \multicolumn{1}{r}{} & \multicolumn{1}{r}{} & \multicolumn{1}{r}{} & \multicolumn{1}{r}{} & \multicolumn{1}{r}{} & \multicolumn{1}{r}{} \\
 & \multicolumn{1}{r}{\nopagebreak $\mu$}  & \multicolumn{1}{r}{$-0.473$} & \multicolumn{1}{r}{$-0.471$} & \multicolumn{1}{r}{$-0.470$} & \multicolumn{1}{r}{$0.387$} & \multicolumn{1}{r}{$0.376$} & \multicolumn{1}{r}{$0.321$} & \multicolumn{1}{r}{$0.533$} & \multicolumn{1}{r}{$0.513$} & \multicolumn{1}{r}{$0.456$} & \multicolumn{1}{r}{} & \multicolumn{1}{r}{} & \multicolumn{1}{r}{} & \multicolumn{1}{r}{} & \multicolumn{1}{r}{} & \multicolumn{1}{r}{} & \multicolumn{1}{r}{} & \multicolumn{1}{r}{} & \multicolumn{1}{r}{} & \multicolumn{1}{r}{} & \multicolumn{1}{r}{} & \multicolumn{1}{r}{} \\
\multicolumn{1}{r}{\rule{0pt}{1.1\normalbaselineskip}\color{lang2}{Haskell}} & \multicolumn{1}{r}{\nopagebreak \color{smedium}{95{\tiny \%}}}  & \multicolumn{1}{r}{\color{lang1}{$-0.512$}} & \multicolumn{1}{r}{\color{lang1}{$-0.502$}} & \multicolumn{1}{r}{\color{lang1}{$-0.494$}} & \multicolumn{1}{r}{\color{lang1}{$-0.011$}} & \multicolumn{1}{r}{$0.000$} & \multicolumn{1}{r}{$0.000$} & \multicolumn{1}{r}{$0.000$} & \multicolumn{1}{r}{$0.000$} & \multicolumn{1}{r}{$0.000$} & \multicolumn{1}{r}{\color{lang1}{$-0.211$}} & \multicolumn{1}{r}{\color{lang1}{$-0.201$}} & \multicolumn{1}{r}{\color{lang1}{$-0.057$}} & \multicolumn{1}{r}{} & \multicolumn{1}{r}{} & \multicolumn{1}{r}{} & \multicolumn{1}{r}{} & \multicolumn{1}{r}{} & \multicolumn{1}{r}{} & \multicolumn{1}{r}{} & \multicolumn{1}{r}{} & \multicolumn{1}{r}{} \\
 & \multicolumn{1}{r}{\nopagebreak \color{sstrong}{99{\tiny \%}}}  & \multicolumn{1}{r}{\color{lang1}{$-0.419$}} & \multicolumn{1}{r}{\color{lang1}{$-0.423$}} & \multicolumn{1}{r}{\color{lang1}{$-0.419$}} & \multicolumn{1}{r}{$+0.000$} & \multicolumn{1}{r}{$+0.000$} & \multicolumn{1}{r}{$+0.000$} & \multicolumn{1}{r}{$+0.000$} & \multicolumn{1}{r}{$+0.000$} & \multicolumn{1}{r}{$+0.000$} & \multicolumn{1}{r}{\color{lang1}{$-0.105$}} & \multicolumn{1}{r}{\color{lang1}{$-0.099$}} & \multicolumn{1}{r}{$+0.000$} & \multicolumn{1}{r}{} & \multicolumn{1}{r}{} & \multicolumn{1}{r}{} & \multicolumn{1}{r}{} & \multicolumn{1}{r}{} & \multicolumn{1}{r}{} & \multicolumn{1}{r}{} & \multicolumn{1}{r}{} & \multicolumn{1}{r}{} \\
 & \multicolumn{1}{r}{\nopagebreak $m$}  & \multicolumn{1}{r}{$-0.687$} & \multicolumn{1}{r}{$-0.681$} & \multicolumn{1}{r}{$-0.653$} & \multicolumn{1}{r}{$-0.318$} & \multicolumn{1}{r}{$-0.306$} & \multicolumn{1}{r}{$+0.055$} & \multicolumn{1}{r}{$+0.009$} & \multicolumn{1}{r}{$+0.001$} & \multicolumn{1}{r}{$+0.229$} & \multicolumn{1}{r}{$-0.443$} & \multicolumn{1}{r}{$-0.429$} & \multicolumn{1}{r}{$-0.243$} & \multicolumn{1}{r}{} & \multicolumn{1}{r}{} & \multicolumn{1}{r}{} & \multicolumn{1}{r}{} & \multicolumn{1}{r}{} & \multicolumn{1}{r}{} & \multicolumn{1}{r}{} & \multicolumn{1}{r}{} & \multicolumn{1}{r}{} \\
 & \multicolumn{1}{r}{\nopagebreak $\mu$}  & \multicolumn{1}{r}{$-0.686$} & \multicolumn{1}{r}{$-0.680$} & \multicolumn{1}{r}{$-0.653$} & \multicolumn{1}{r}{$-0.317$} & \multicolumn{1}{r}{$-0.305$} & \multicolumn{1}{r}{$0.057$} & \multicolumn{1}{r}{$0.007$} & \multicolumn{1}{r}{$0.001$} & \multicolumn{1}{r}{$0.233$} & \multicolumn{1}{r}{$-0.441$} & \multicolumn{1}{r}{$-0.427$} & \multicolumn{1}{r}{$-0.242$} & \multicolumn{1}{r}{} & \multicolumn{1}{r}{} & \multicolumn{1}{r}{} & \multicolumn{1}{r}{} & \multicolumn{1}{r}{} & \multicolumn{1}{r}{} & \multicolumn{1}{r}{} & \multicolumn{1}{r}{} & \multicolumn{1}{r}{} \\
\multicolumn{1}{r}{\rule{0pt}{1.1\normalbaselineskip}\color{lang2}{Java}} & \multicolumn{1}{r}{\nopagebreak \color{smedium}{95{\tiny \%}}}  & \multicolumn{1}{r}{\color{lang1}{$-0.461$}} & \multicolumn{1}{r}{\color{lang1}{$-0.457$}} & \multicolumn{1}{r}{\color{lang1}{$-0.575$}} & \multicolumn{1}{r}{$0.000$} & \multicolumn{1}{r}{$0.000$} & \multicolumn{1}{r}{$0.000$} & \multicolumn{1}{r}{\color{lang2}{$0.280$}} & \multicolumn{1}{r}{\color{lang2}{$0.278$}} & \multicolumn{1}{r}{\color{lang2}{$0.240$}} & \multicolumn{1}{r}{$0.000$} & \multicolumn{1}{r}{$0.000$} & \multicolumn{1}{r}{\color{lang1}{$-0.011$}} & \multicolumn{1}{r}{$0.000$} & \multicolumn{1}{r}{$0.000$} & \multicolumn{1}{r}{$0.000$} & \multicolumn{1}{r}{} & \multicolumn{1}{r}{} & \multicolumn{1}{r}{} & \multicolumn{1}{r}{} & \multicolumn{1}{r}{} & \multicolumn{1}{r}{} \\
 & \multicolumn{1}{r}{\nopagebreak \color{sstrong}{99{\tiny \%}}}  & \multicolumn{1}{r}{\color{lang1}{$-0.397$}} & \multicolumn{1}{r}{\color{lang1}{$-0.389$}} & \multicolumn{1}{r}{\color{lang1}{$-0.552$}} & \multicolumn{1}{r}{$+0.000$} & \multicolumn{1}{r}{$+0.000$} & \multicolumn{1}{r}{$+0.000$} & \multicolumn{1}{r}{\color{lang2}{$+0.186$}} & \multicolumn{1}{r}{\color{lang2}{$+0.182$}} & \multicolumn{1}{r}{\color{lang2}{$+0.173$}} & \multicolumn{1}{r}{$+0.000$} & \multicolumn{1}{r}{$+0.000$} & \multicolumn{1}{r}{$+0.000$} & \multicolumn{1}{r}{$+0.000$} & \multicolumn{1}{r}{$+0.000$} & \multicolumn{1}{r}{$+0.000$} & \multicolumn{1}{r}{} & \multicolumn{1}{r}{} & \multicolumn{1}{r}{} & \multicolumn{1}{r}{} & \multicolumn{1}{r}{} & \multicolumn{1}{r}{} \\
 & \multicolumn{1}{r}{\nopagebreak $m$}  & \multicolumn{1}{r}{$-0.610$} & \multicolumn{1}{r}{$-0.606$} & \multicolumn{1}{r}{$-0.632$} & \multicolumn{1}{r}{$+0.187$} & \multicolumn{1}{r}{$+0.177$} & \multicolumn{1}{r}{$+0.121$} & \multicolumn{1}{r}{$+0.520$} & \multicolumn{1}{r}{$+0.511$} & \multicolumn{1}{r}{$+0.447$} & \multicolumn{1}{r}{$-0.193$} & \multicolumn{1}{r}{$-0.190$} & \multicolumn{1}{r}{$-0.187$} & \multicolumn{1}{r}{$+0.217$} & \multicolumn{1}{r}{$+0.216$} & \multicolumn{1}{r}{$+0.123$} & \multicolumn{1}{r}{} & \multicolumn{1}{r}{} & \multicolumn{1}{r}{} & \multicolumn{1}{r}{} & \multicolumn{1}{r}{} & \multicolumn{1}{r}{} \\
 & \multicolumn{1}{r}{\nopagebreak $\mu$}  & \multicolumn{1}{r}{$-0.610$} & \multicolumn{1}{r}{$-0.606$} & \multicolumn{1}{r}{$-0.632$} & \multicolumn{1}{r}{$0.189$} & \multicolumn{1}{r}{$0.177$} & \multicolumn{1}{r}{$0.120$} & \multicolumn{1}{r}{$0.520$} & \multicolumn{1}{r}{$0.511$} & \multicolumn{1}{r}{$0.443$} & \multicolumn{1}{r}{$-0.193$} & \multicolumn{1}{r}{$-0.190$} & \multicolumn{1}{r}{$-0.186$} & \multicolumn{1}{r}{$0.219$} & \multicolumn{1}{r}{$0.216$} & \multicolumn{1}{r}{$0.120$} & \multicolumn{1}{r}{} & \multicolumn{1}{r}{} & \multicolumn{1}{r}{} & \multicolumn{1}{r}{} & \multicolumn{1}{r}{} & \multicolumn{1}{r}{} \\
\multicolumn{1}{r}{\rule{0pt}{1.1\normalbaselineskip}\color{lang2}{Python}} & \multicolumn{1}{r}{\nopagebreak \color{smedium}{95{\tiny \%}}}  & \multicolumn{1}{r}{\color{lang1}{$-0.581$}} & \multicolumn{1}{r}{\color{lang1}{$-0.569$}} & \multicolumn{1}{r}{\color{lang1}{$-0.791$}} & \multicolumn{1}{r}{\color{lang1}{$-0.102$}} & \multicolumn{1}{r}{\color{lang1}{$-0.103$}} & \multicolumn{1}{r}{\color{lang1}{$-0.164$}} & \multicolumn{1}{r}{$0.000$} & \multicolumn{1}{r}{$0.000$} & \multicolumn{1}{r}{$0.000$} & \multicolumn{1}{r}{\color{lang1}{$-0.232$}} & \multicolumn{1}{r}{\color{lang1}{$-0.224$}} & \multicolumn{1}{r}{\color{lang1}{$-0.308$}} & \multicolumn{1}{r}{$0.000$} & \multicolumn{1}{r}{$0.000$} & \multicolumn{1}{r}{$0.000$} & \multicolumn{1}{r}{\color{lang1}{$-0.083$}} & \multicolumn{1}{r}{\color{lang1}{$-0.080$}} & \multicolumn{1}{r}{\color{lang1}{$-0.161$}} & \multicolumn{1}{r}{} & \multicolumn{1}{r}{} & \multicolumn{1}{r}{} \\
 & \multicolumn{1}{r}{\nopagebreak \color{sstrong}{99{\tiny \%}}}  & \multicolumn{1}{r}{\color{lang1}{$-0.515$}} & \multicolumn{1}{r}{\color{lang1}{$-0.502$}} & \multicolumn{1}{r}{\color{lang1}{$-0.765$}} & \multicolumn{1}{r}{$+0.000$} & \multicolumn{1}{r}{$+0.000$} & \multicolumn{1}{r}{\color{lang1}{$-0.062$}} & \multicolumn{1}{r}{$+0.000$} & \multicolumn{1}{r}{$+0.000$} & \multicolumn{1}{r}{$+0.000$} & \multicolumn{1}{r}{\color{lang1}{$-0.138$}} & \multicolumn{1}{r}{\color{lang1}{$-0.125$}} & \multicolumn{1}{r}{\color{lang1}{$-0.224$}} & \multicolumn{1}{r}{$+0.000$} & \multicolumn{1}{r}{$+0.000$} & \multicolumn{1}{r}{$+0.000$} & \multicolumn{1}{r}{$+0.000$} & \multicolumn{1}{r}{$+0.000$} & \multicolumn{1}{r}{\color{lang1}{$-0.077$}} & \multicolumn{1}{r}{} & \multicolumn{1}{r}{} & \multicolumn{1}{r}{} \\
 & \multicolumn{1}{r}{\nopagebreak $m$}  & \multicolumn{1}{r}{$-0.728$} & \multicolumn{1}{r}{$-0.724$} & \multicolumn{1}{r}{$-0.851$} & \multicolumn{1}{r}{$-0.375$} & \multicolumn{1}{r}{$-0.371$} & \multicolumn{1}{r}{$-0.404$} & \multicolumn{1}{r}{$-0.021$} & \multicolumn{1}{r}{$-0.017$} & \multicolumn{1}{r}{$-0.150$} & \multicolumn{1}{r}{$-0.445$} & \multicolumn{1}{r}{$-0.436$} & \multicolumn{1}{r}{$-0.495$} & \multicolumn{1}{r}{$+0.003$} & \multicolumn{1}{r}{$+0.004$} & \multicolumn{1}{r}{$-0.117$} & \multicolumn{1}{r}{$-0.319$} & \multicolumn{1}{r}{$-0.316$} & \multicolumn{1}{r}{$-0.369$} & \multicolumn{1}{r}{} & \multicolumn{1}{r}{} & \multicolumn{1}{r}{} \\
 & \multicolumn{1}{r}{\nopagebreak $\mu$}  & \multicolumn{1}{r}{$-0.728$} & \multicolumn{1}{r}{$-0.723$} & \multicolumn{1}{r}{$-0.851$} & \multicolumn{1}{r}{$-0.376$} & \multicolumn{1}{r}{$-0.371$} & \multicolumn{1}{r}{$-0.405$} & \multicolumn{1}{r}{$-0.023$} & \multicolumn{1}{r}{$-0.016$} & \multicolumn{1}{r}{$-0.158$} & \multicolumn{1}{r}{$-0.445$} & \multicolumn{1}{r}{$-0.437$} & \multicolumn{1}{r}{$-0.497$} & \multicolumn{1}{r}{$0.004$} & \multicolumn{1}{r}{$0.004$} & \multicolumn{1}{r}{$-0.119$} & \multicolumn{1}{r}{$-0.319$} & \multicolumn{1}{r}{$-0.315$} & \multicolumn{1}{r}{$-0.372$} & \multicolumn{1}{r}{} & \multicolumn{1}{r}{} & \multicolumn{1}{r}{} \\
\multicolumn{1}{r}{\rule{0pt}{1.1\normalbaselineskip}\color{lang2}{Ruby}} & \multicolumn{1}{r}{\nopagebreak \color{smedium}{95{\tiny \%}}}  & \multicolumn{1}{r}{\color{lang1}{$-0.452$}} & \multicolumn{1}{r}{\color{lang1}{$-0.443$}} & \multicolumn{1}{r}{\color{lang1}{$-0.950$}} & \multicolumn{1}{r}{\color{lang1}{$-0.288$}} & \multicolumn{1}{r}{\color{lang1}{$-0.274$}} & \multicolumn{1}{r}{\color{lang1}{$-0.548$}} & \multicolumn{1}{r}{$0.000$} & \multicolumn{1}{r}{$0.000$} & \multicolumn{1}{r}{\color{lang1}{$-0.568$}} & \multicolumn{1}{r}{\color{lang1}{$-0.321$}} & \multicolumn{1}{r}{\color{lang1}{$-0.312$}} & \multicolumn{1}{r}{\color{lang1}{$-0.631$}} & \multicolumn{1}{r}{$0.000$} & \multicolumn{1}{r}{$0.000$} & \multicolumn{1}{r}{\color{lang1}{$-0.858$}} & \multicolumn{1}{r}{\color{lang1}{$-0.255$}} & \multicolumn{1}{r}{\color{lang1}{$-0.244$}} & \multicolumn{1}{r}{\color{lang1}{$-0.807$}} & \multicolumn{1}{r}{$0.000$} & \multicolumn{1}{r}{$0.000$} & \multicolumn{1}{r}{\color{lang1}{$-0.008$}} \\
 & \multicolumn{1}{r}{\nopagebreak \color{sstrong}{99{\tiny \%}}}  & \multicolumn{1}{r}{\color{lang1}{$-0.366$}} & \multicolumn{1}{r}{\color{lang1}{$-0.347$}} & \multicolumn{1}{r}{\color{lang1}{$-0.947$}} & \multicolumn{1}{r}{\color{lang1}{$-0.163$}} & \multicolumn{1}{r}{\color{lang1}{$-0.150$}} & \multicolumn{1}{r}{\color{lang1}{$-0.488$}} & \multicolumn{1}{r}{$+0.000$} & \multicolumn{1}{r}{$+0.000$} & \multicolumn{1}{r}{\color{lang1}{$-0.505$}} & \multicolumn{1}{r}{\color{lang1}{$-0.222$}} & \multicolumn{1}{r}{\color{lang1}{$-0.211$}} & \multicolumn{1}{r}{\color{lang1}{$-0.575$}} & \multicolumn{1}{r}{$+0.000$} & \multicolumn{1}{r}{$+0.000$} & \multicolumn{1}{r}{\color{lang1}{$-0.843$}} & \multicolumn{1}{r}{\color{lang1}{$-0.161$}} & \multicolumn{1}{r}{\color{lang1}{$-0.154$}} & \multicolumn{1}{r}{\color{lang1}{$-0.786$}} & \multicolumn{1}{r}{$+0.000$} & \multicolumn{1}{r}{$+0.000$} & \multicolumn{1}{r}{$+0.000$} \\
 & \multicolumn{1}{r}{\nopagebreak $m$}  & \multicolumn{1}{r}{$-0.656$} & \multicolumn{1}{r}{$-0.646$} & \multicolumn{1}{r}{$-0.957$} & \multicolumn{1}{r}{$-0.545$} & \multicolumn{1}{r}{$-0.529$} & \multicolumn{1}{r}{$-0.699$} & \multicolumn{1}{r}{$-0.026$} & \multicolumn{1}{r}{$-0.026$} & \multicolumn{1}{r}{$-0.723$} & \multicolumn{1}{r}{$-0.542$} & \multicolumn{1}{r}{$-0.538$} & \multicolumn{1}{r}{$-0.749$} & \multicolumn{1}{r}{$-0.114$} & \multicolumn{1}{r}{$-0.115$} & \multicolumn{1}{r}{$-0.897$} & \multicolumn{1}{r}{$-0.470$} & \multicolumn{1}{r}{$-0.460$} & \multicolumn{1}{r}{$-0.854$} & \multicolumn{1}{r}{$-0.173$} & \multicolumn{1}{r}{$-0.172$} & \multicolumn{1}{r}{$-0.175$} \\
 & \multicolumn{1}{r}{\nopagebreak $\mu$}  & \multicolumn{1}{r}{$-0.655$} & \multicolumn{1}{r}{$-0.644$} & \multicolumn{1}{r}{$-0.957$} & \multicolumn{1}{r}{$-0.543$} & \multicolumn{1}{r}{$-0.527$} & \multicolumn{1}{r}{$-0.698$} & \multicolumn{1}{r}{$-0.029$} & \multicolumn{1}{r}{$-0.024$} & \multicolumn{1}{r}{$-0.723$} & \multicolumn{1}{r}{$-0.542$} & \multicolumn{1}{r}{$-0.536$} & \multicolumn{1}{r}{$-0.749$} & \multicolumn{1}{r}{$-0.113$} & \multicolumn{1}{r}{$-0.114$} & \multicolumn{1}{r}{$-0.897$} & \multicolumn{1}{r}{$-0.469$} & \multicolumn{1}{r}{$-0.460$} & \multicolumn{1}{r}{$-0.854$} & \multicolumn{1}{r}{$-0.172$} & \multicolumn{1}{r}{$-0.171$} & \multicolumn{1}{r}{$-0.175$} \\
\bottomrule 
\end{tabular}

\caption{
Bayesian reanalysis of \rosetta's Rosetta Code data about the performance of 8 programming languages.
For each comparison of language $\ell_1$ ({\color{lang1} column header}) with language $\ell_2$ ({\color{lang2}row header}),
  for each choice of prior distribution among uniform \up, centered normal \np, and shifted normal \spd,
  the table reports 
  the endpoint of the 95\% and 99\% uncertainty intervals of the posterior inverse speedup of $\ell_1$ vs.~$\ell_2$ that is closest to the origin,
  if such interval does \emph{not} include the origin;
  in this case, the endpoint's absolute value is a lower bound on the inverse speedup of $\ell_1$ vs.~$\ell_2$,
  and indicates that $\ell_1$ tends to be faster if it is {\color{lang1} negative} and that $\ell_2$ tends to be faster if it is {\color{lang2} positive}.
  If the uncertainty interval includes the origin, the table reports a value of 0.0, which indicates the performance comparison is inconclusive.
  The table also reports median $m$ and mean $\mu$ of the \emph{posterior} inverse speedup of $\ell_1$ vs.~$\ell_2$ (again, negative values indicate that $\ell_1$ is faster on average, and positive values that $\ell_2$ is).}
\label{tab:bayes_cAR}
\end{sidewaystable*}

\begin{figure*}[!tb]
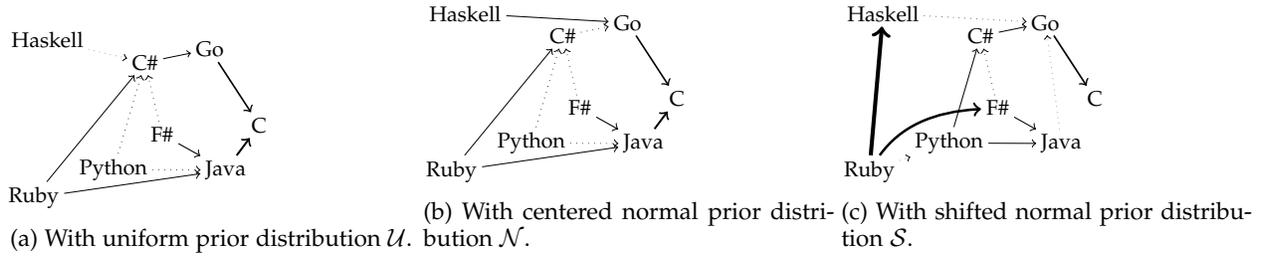

  \centering
  \begin{subfigure}[b]{0.3\linewidth}
    \rosettagraph{graph_bayes_u_cAR}
    \caption{With uniform prior distribution \up.}
    \label{fig:bayes_u_cAR}
  \end{subfigure}
  \begin{subfigure}[b]{0.3\linewidth}
    \rosettagraph{graph_bayes_nBcAR}
    \caption{With centered normal prior distribution \np.}
    \label{fig:bayes_nBcAR}
  \end{subfigure}
  \begin{subfigure}[b]{0.3\linewidth}
    \rosettagraph{graph_bayes_sBcAR}    
    \caption{With shifted normal prior distribution \spd.}
    \label{fig:bayes_sBcAR}
  \end{subfigure}
\caption{Language relationship graphs summarizing the Bayesian reanalysis of \rosetta, with different prior distributions of speedups.
  In every graph, an arrow from node $\ell_1$ to node $\ell_2$ indicates that, according to the posterior probability distribution of inverse speedups,
  $\ell_1$ is faster than $\ell_2$ with 99\% probability (solid line) or 95\% probability (dotted line);
  the thickness of the arrows is proportional to the absolute value of the minimum inverse speedup.}
  \label{fig:bayes_R}
\end{figure*}

\begin{table}[!h]
  \setlength{\tabcolsep}{2pt}
  \centering
  \scriptsize
  \begin{tabular}{lcccccccccccc}
\toprule
 & \multicolumn{8}{c}{$\mathcal{U}$} & \multicolumn{2}{c}{$\mathcal{N}$} & \multicolumn{2}{c}{$\mathcal{S}$} \\ \cmidrule(lr){2-9}\cmidrule(lr){10-11}\cmidrule(lr){12-13}
 & \multicolumn{2}{c}{--} & \multicolumn{2}{c}{B} & \multicolumn{2}{c}{H} & \multicolumn{2}{c}{BH} & \multicolumn{2}{c}{--} & \multicolumn{2}{c}{--} \\ \cmidrule(lr){2-3}\cmidrule(lr){4-5}\cmidrule(lr){6-7}\cmidrule(lr){8-9}\cmidrule(lr){10-11}\cmidrule(lr){12-13}
\textsc{analysis}  & $95\,\%$ & $99\,\%$ & $95\,\%$ & $99\,\%$ & $95\,\%$ & $99\,\%$ & $95\,\%$ & $99\,\%$ & $95\,\%$ & $99\,\%$ & $95\,\%$ & \multicolumn{1}{c}{$99\,\%$} \\ 
\midrule
\nopagebreak frequentist  & $17$ & $12$ & $7$ & $6$ & $7$ & $6$ & $15$ & $7$ &  &  &  &  \\
\rule{0pt}{1.2\normalbaselineskip}Bayesian  & $19$ & $15$ &  &  &  &  &  &  & $18$ & $14$ & $22$ & $18$ \\
\bottomrule 
\end{tabular}

  \caption{The number of language comparisons, out of a total 28 language pairs, where each \textsc{analysis} found a significant difference
    at the 95\% and 99\% levels.
    The two rows correspond to \emph{frequentist} and \emph{Bayesian} analyses;
    the frequentist analyses differ in what kind of $p$-value correction they employ (-- for none, B for Bonferroni, H for Holm, and BH for Benjamini-Hochberg);
    the Bayesian analyses differ in what prior they employ (\up{} for uniform, \np {}for centered normal, and \spd{} for shifted normal distributions).
  }
  \label{tab:significance_comparison}
\end{table}

\subsubsection{Bayesian Reanalysis Results}
\label{sec:rosetta:bayes-results}
\autoref{tab:bayes_cAR} shows the results of the Bayesian analysis of \rosetta data, for each of three priors.
If two languages differ at a {\color{sweak} 95\%} certainty level, the table reports the value $c_{95\%}$ if the language in the {\color{lang1} column} header is faster, and the value $c_{5\%}$ if the language in the {\color{lang2} row} header is faster; if the comparison is inconclusive, the table reports the value $0.0$.
The table displays the results of the comparison at a {\color{sstrong} 99\%} certainty level in a similar way, using the values $c_{99\%}$, $c_{1\%}$, or $0.0$.

The \emph{language relationship graphs} in \autoref{fig:bayes_R} summarize all comparisons similarly to the frequentist analysis: nodes are languages; an arrow from $\ell_1$ to $\ell_2$ denotes that the difference between the two languages is significant (at 95\% certainty level), and goes from the slower to the faster language; 
solid arrows denote differences that are significant at 99\% certainty level (otherwise arrows are dotted);
the thickness of the arrows is proportional to the absolute value of the endpoint of the 95\% or 99\% certainty level closer to the origin (as explained above).

\myparagraph{Prediction errors and multiple tests.}
Multiple testing is generally not an issue in Bayesian analysis~\cite{MultipleNoWorry}.
In fact, there is no null-hypothesis testing in Bayesian analysis:
we are not performing multiple tests, each with a binary outcome,
but are computing posterior distribution probabilities based on data and priors.
We can estimate mean and standard deviation of the posterior,
and get a clear, quantitative picture of how likely there is an effect and of what size---instead
of having to guess a correction for how the errors in binary tests may compound.
As we discussed in \autoref{sec:classic-vs-bayes}, it is also advisable to focus on type $S$ and $M$ errors,
which are directly meaningful in terms of practical significance.

While it is possible to reason in terms of type $S$ and $M$ errors also using frequentist techniques such as effect sizes,
Bayesian analysis provide a natural framework to precisely and soundly assess these probabilities of errors.
In our Bayesian reanalysis of \rosetta's data, we can readily assess the probability of type $S$ and $M$ errors:
\begin{itemize}
\item if a 95\% uncertainty interval does not include the origin, there is at most a 5\% chance of committing a type $S$ error in that comparison; \item the width $W$ of a 95\% uncertainty interval indicates that there is at most a 5\% chance of committing a type $M$ error larger than $W$. \end{itemize}

\myparagraph{Significance analysis.}
There are no major inconsistencies between the overall pictures drawn by the frequentist (with effect sizes) and by the Bayesian analysis. Bayesian statistics, however, produced results that are more natural to interpret in terms of practically significant measures,
and lend richer information that supports a detailed analysis
even when the frequentist analysis was utterly inconclusive.

In quantitative terms, every ``significant'' difference according to frequentist analysis was confirmed by Bayesian analysis.
Conversely, the Bayesian analysis provides conclusive comparisons (that is, one language being faster than another) for more language pairs than the frequentist analysis (see \autoref{tab:significance_comparison}):
while the latter finds significant differences in 17 language pairs (out of a total of 28),
Bayesian analysis finds 18 with centered normal priors, 19 with uniform priors, and 22 with shifted normal priors.

We consider this a distinct advantage from a software engineering perspective: we could draw more conclusions from the same data. The fact that priors of different strengths lead to increasingly precise outcomes also sharpens the overall conclusions:
the Rosetta dataset and the Bench dataset (used to model the stronger prior) seem to often corroborate each other.
More generally, such techniques support analyses where each new scientific study increases the clarity and robustness of the overall understanding of what is really going on.
Even in situations where a new study does not corroborate but contradicts a previous one, we would enrich our understanding of the issues by making it more nuanced in a quantitative way.

\subsubsection{Prior Sensitivity Analysis}
\label{sec:rosetta:bayes:sensitivity}
Differences due to using different priors point to what kinds of assumptions are necessary to narrow down the analysis in certain cases.
To disentangle the cases where weak data leads to inconclusive or inconsistent results, we can always visually and numerically inspect the \emph{posterior distributions} provided by Bayesian analysis.
This is a kind of \emph{sensitivity analysis} of the priors, whose goal is reporting the effects of choosing different priors,
in order to better understand the strength of the data under analysis and the ramifications of assumptions that we may introduce.

\begin{figure*}[!htb]
  \centering
  \begin{subfigure}[b]{0.4\textwidth}
    \resizebox{1.0\textwidth}{!}{\begin{tikzpicture}\end{tikzpicture}}
  \end{subfigure}
  \begin{subfigure}[b]{0.4\textwidth}
    \resizebox{1.0\textwidth}{!}{\begin{tikzpicture}\end{tikzpicture}}
  \end{subfigure}\\[\abovecaptionskip]
  \begin{subfigure}[b]{0.4\textwidth}
    \resizebox{1.0\textwidth}{!}{\begin{tikzpicture}\end{tikzpicture}}
  \end{subfigure}
  \begin{subfigure}[b]{0.4\textwidth}
    \resizebox{1.0\textwidth}{!}{\begin{tikzpicture}\end{tikzpicture}}
  \end{subfigure}
  \caption{Posterior distributions of inverse speedup ratios comparing C\# to F\#, F\# to Python, F\# to Ruby, and Haskell to Python.
  Each comparison comprises three posterior distributions built using Bayesian analysis with uniform $\up$, centered normal $\np$, and shifted normal $\spd$ prior distributions.}
  \label{fig:posteriors-fsharp-vs-three}
\end{figure*}
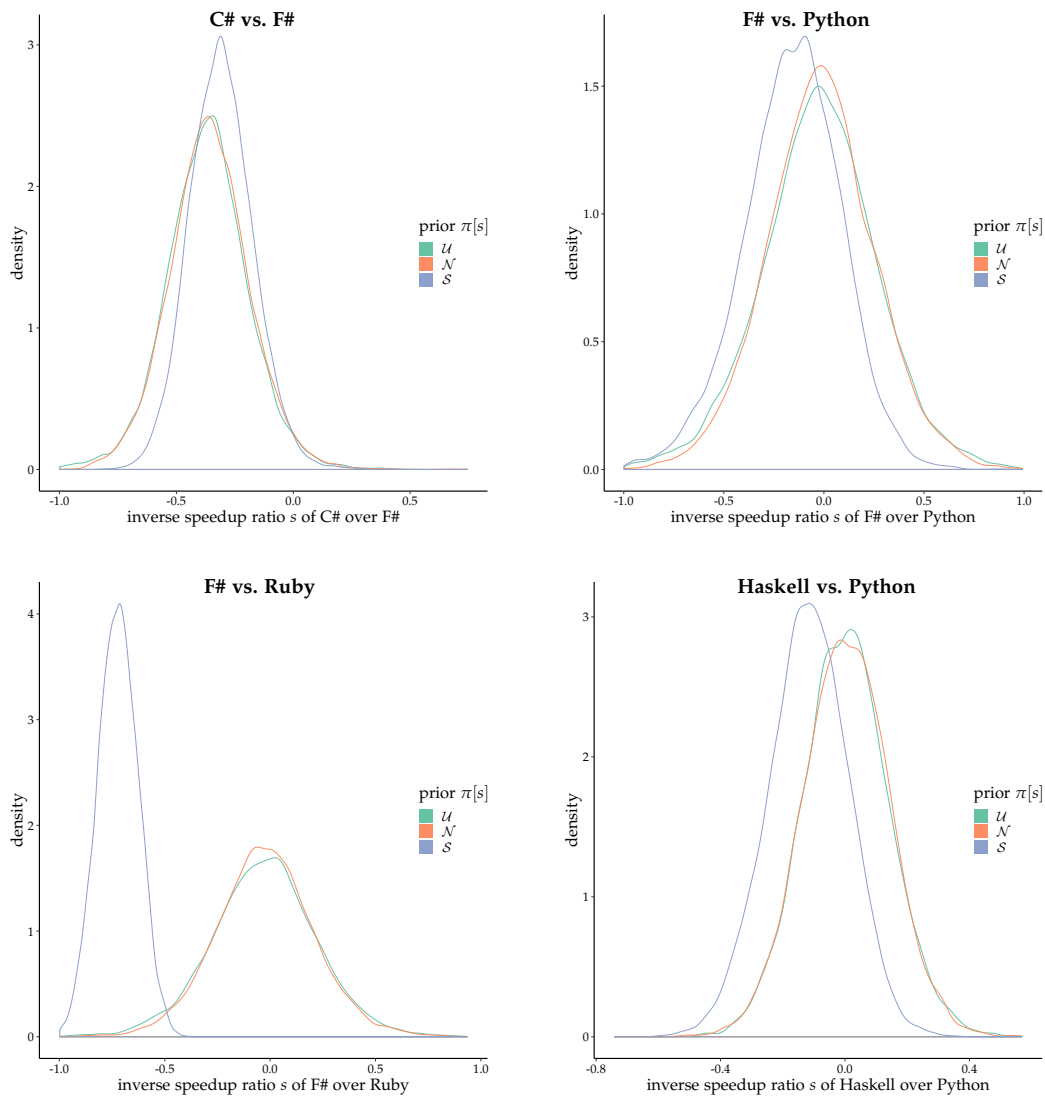

Rather than going through an exhaustive sensitivity analysis, let us single out four interesting cases.
These are the four language comparisons that we mentioned at the end of \autoref{sec:rosetta:summary-bayes}:
C\# vs.\ F\#, F\# vs.\ Python, F\# vs.\ Ruby, and Haskell vs.\ Python.
These comparisons were inconclusive in the frequentist analysis, sometimes with contradictory evidence given by different summary statistics.
In the Bayesian analysis, we get nuanced yet clearer results;
and we can go beyond a binary choice made on the basis of summary statistics, looking at the posterior distributions plot for these three cases
(shown in \autoref{fig:posteriors-fsharp-vs-three}):
\begin{description}
\item[C\# vs.\ F\#:]
  regardless of the prior, the posterior estimated (inverse) speedup is negative 
  with 95\% probability.
  This confirms that C\# tends to be faster in the experiments, largely independent of prior assumptions: the data \emph{swamps} the priors.
  
\item[F\# vs.\ Python:]
  the analysis is genuinely inconclusive if we use an unbiased prior (uniform or centered normal), as both posteriors are basically centered around the origin.
  If we use a biased prior (the shifted normal prior), instead, the posterior tends to negative values.
  This may indicate some evidence that F\# is faster,
  but only if we are willing to use \bench's data as the starting point to process Rosetta Code's data
  (which, alone, remains too weak to suggest a significant effect).

  Precisely, we can compute on the posterior draws---built using the shifted prior---that
  the probability that the speedup is negative is approximately 73\%; and the probability that it is less than $-0.09$ (corresponding to a $1.1$-or-greater speedup of F\# over Python) is about 60\%.
  The shifted normal prior is $\np(-0.55, 0.69)$ in this case, which corresponds to a marked bias in favor of F\# being substantially faster.
  Nonetheless, we can find both tasks where F\# is faster and tasks where Python is faster
  both in Rosetta Code and in \bench, indicating no clear winner with the current data even if F\# has some advantage on average.

\item[F\# vs.\ Ruby:]
  here the posterior changes drastically according to the prior.
  With uniform and centered normal priors, we cannot claim any noticeable speedup of one language over the other.
  But with the shifted normal prior, the posterior lies decidedly to the left of the origin and is quite narrow,
  indicating a high probability that F\# is much faster than Ruby;
  precisely, the speedup is less than $-0.4$ with near certainty.

  In this comparison, the shifted prior is $\np(-0.79,  0.31)$---strongly biased in favor of F\#.
  The dependence on such a strong prior means that Rosetta's data about this language comparison is not conclusive by its own:
  both datasets may indicate that F\# tends to be faster, but \bench's data is overwhelming whereas Rosetta's is very mixed.
  Regardless of whether we are willing to take \bench as a starting point of our analysis,
  we understand that Rosetta's data about this language comparison is inadequate and should be extended and sharpened with new experiments
  or with a deeper analysis of Rosetta's experimental conditions that determined the outlier results.

\item[Haskell vs.\ Python:]
  the language comparison is inconclusive if we use an unbiased prior (uniform or centered normal).   In contrast, starting from a shifted normal prior, Haskell emerges as the faster language with 82\% probability.

  In this case the biased prior is $\np(-0.67, 0.55)$, which is clearly biased in favor of Haskell but would not prevent strong data evidence from changing the outcome.
  If we are willing to see this biased prior as encoding a reasonable assumption on the preferential advantage that a compiled language such as Haskell
  ought to have over an interpreted language, Rosetta's data tempers the speed advantage Haskell can claim but does not nullify it.
  Regardless of whether we find this interpretation reasonable, the analysis clearly shows what the outcome of this language comparison hinges on,
  and suggests that further experiments involving these two languages are needed to get to a more general conclusion.
\end{description}

Repeating the sensitivity analysis for every language comparison, extending the models, and possibly collecting more data,
are beyond the scope of this paper.
The goal of this subsection was illustrating concrete examples of how sensitivity analysis can support a sophisticated data analysis
while remaining grounded in results that have a direct connection to the quantities of interest that determine practical significance.
Key to this flexibility are the very features of Bayesian statistics that we presented throughout the paper.

\subsubsection{Summary of Bayesian Statistics}
\label{sec:rosetta:summary-bayes}
The availability of full posterior distributions makes Bayesian analyses more transparent
and based on richer, multi-dimensional information than frequentist analyses.
Priors are another tool unique to Bayesian analysis that fully supports ``what-if'' analyses:
by trying out different priors---from completely uniform to informative/biased---we can 
understand the impact of assumptions on interpreting the experimental data,
and naturally connect our analysis to others and, more generally, to the body of knowledge
of the domain under empirical investigation.

\section{Doing Bayesian Statistics in Practice: Guidelines and Tools}
\label{sec:guidelines}

The main goal of this paper is demonstrating the advantages of Bayesian statistics over frequentist statistics.
Once we become convinced of this point, how do we \emph{concretely} adopt Bayesian statistics in the kind of empirical research that is common in software engineering?
While we leave a detailed presentation of a framework to future work, here we summarize some basic guidelines organized around a few broadly applicable steps.

\myparagraph{Research questions.}
If we formulate the research questions following the usual approach based on null hypothesis (no effect) vs.~alternative hypothesis (some effect),
then using statistical hypothesis testing may appear as the only natural analysis approach.
To avoid this, we should try to phrase the research questions in a way that:
\begin{enumerate*}
\item they have a clear connection to variables in the \emph{analysis domain};
\item they \emph{avoid} dichotomous alternative answers; and
\item they \emph{quantify}---if approximately---the expected ranges of values that constitute a \emph{practically} significant effect.
\end{enumerate*}

For example, we could rephrase \rosetta's generic research question about two languages $\ell_1$ and $\ell_2$:
\begin{quotation}
  Is $\ell_1$ significantly faster than $\ell_2$?
\end{quotation}
as:
\begin{quotation}
  What is the average speedup of $\ell_1$ over $\ell_2$,
  and what is the probability that it is greater than 1\%?\footnote{In \autoref{sec:bayesian-rosetta}, we answered a different question---about the probability that it is greater than zero---so as to allow a direct comparison to the results of the original analysis based on hypothesis testing.}
\end{quotation}
The new formulation requires that every answer to the question provide an estimate of speedup,
and that small (less than 1\%) differences are practically irrelevant.

\myparagraph{Regressive models.}
Regressive models---from simple linear regressions to more flexible generalized linear models---are very powerful statistical models
which are applicable with remarkable flexibility to a wide variety of problems and data domains~\cite{GelmanHill-linearmodels}.\footnote{In particular, regressive models are not limited to continuous variables but can also model discrete variables such as those on a Likert scale~\cite{bayes-likert}.} Since regression can also be done by means of frequentist techniques, regressive models are an approachable entry point to Bayesian statistics for researcher used to frequentist techniques. There exist statistical packages (such as \verb|rstanarm| and \verb|brms| for R) that provide functions with an interface very similar to the classic \verb|lm| (which does regression with frequentist techniques), with optional  additional inputs (for example, to change the priors) to take full advantage of Bayesian statistics' greater flexibility.
We used such packages in the reanalysis of \autotest{} in \autoref{sec:testing}.

Even when a statistical model other than standard linear regression is a better choice, it often is useful to start the analysis using simple regressive models, and then tweak them with the additional modeling features that may be required.
This incrementality may not be possible---or as easy---in a frequentist setting, where changing statistical model normally
requires switching to a new custom algorithm.
In contrast, Bayesian statistics are all based on a uniform framework that applies Bayes' theorem to update the posterior based on the data;
the model is all in the choice of priors and likelihoods.
Thus, performing software engineering studies based on Bayesian techniques would support a flexible use of different models of varying complexity within a single, coherent statistical framework.

\myparagraph{Choosing priors.}
After following this paper's reanalyses, it should be clear that ``choosing the prior'' need not be a complex exercise.
In fact, in Bayesian analysis you don't \emph{choose the} prior, but rather experiment with a variety of priors as a way of evaluating different modeling choices.
This process, presented in \autoref{sec:role-of-priors}, is called \emph{sensitivity analysis}.

As a minimum, priors should indicate reasonable ranges of values, or, conversely, rule out ``impossible'' ranges of values as highly unlikely.
This is what the uniform and centered normal priors did in \autoref{sec:bayesian-rosetta}.
If we have more precise information about the plausible variable ranges and we want to explore their effect on the model, we can try out more biased priors---such as the shifted normal priors in  \autoref{sec:bayesian-rosetta}.
The goal of sensitivity analysis is not to choose a prior over others---not even after the analysis---but
rather to explore the same data from different angles, painting a rich, nuanced picture.

When the goal of sensitivity analysis is to identify the model with the greatest \emph{predictive power}---that is, that strikes a good balance between underfitting and overfitting---there
exist quantitative comparison criteria based on measures of information entropy~\cite{rethinking}. In most common cases, however, a qualitative sensitivity analysis---such as the one in \autoref{sec:rosetta:bayes:sensitivity}---goes a long way
in describing the analysis results while remaining grounded in the problem domain.

As more empirical evidence is gathered, and studies are performed, about a certain software engineering area of interest, priors can be chosen based on the evidence that is already known and is relevant to new studies within the same area.
This can give empirical software engineering a concrete way in which to build chains of evidence that span multiple studies, and ultimately present the current state-of-the-art knowledge to practitioners in a quantitative form. Doing so could also make research results more actionable, since different scenarios could be evaluated and simulated.

\myparagraph{Visualization.}
At every step during the analysis, visualizing data and intermediate results acts as a sanity check on the data analysis.
Of course visualization is not helpful only in a Bayesian setting, but one of the key strengths of Bayesian analysis is that it often better lends itself to visualization since it is based on \emph{distributions of data}, and how they change to better reflect the data.
For example, the posterior plots of \autoref{fig:posteriors-fsharp-vs-three} support the sensitivity analysis of three language comparisons that closes \autoref{sec:bayesian-rosetta}.\footnote{Gabry et al.~\cite{gabry2019visualization} suggest guidelines on visualization in a Bayesian analysis workflow.}

\myparagraph{Dichotomous decisions.}
Even if we avoid using the problematic notion of ``null hypothesis'', there exist research questions that ultimately require a dichotomous (yes/no) answer.
How can we approach such questions without giving up the benefits of a Bayesian analysis?

Since Bayesian analysis's primary source of information is in the form of posterior distributions,
the best way to approach dichotomous decisions is to ground them in an \emph{analysis of the posteriors}.
This also helps ensure that the decisions are taken based on \emph{practically significant} measures rather than artificial statistics
such as the $p$-values.

This general guideline is very flexible.
Then, there exist more specific protocols that can be applied to formalize the notion of how to answer dichotomous questions based on an analysis of the posteriors. In clinical studies, for instance, researchers have used the notion of \emph{region of practical equivalence} (ROPE) to express ``a range of parameter values that are equivalent to the null value for current practical purpose''~\cite{bayesianNewStats-ROPE}.
ROPEs are typically derived based on historical data and meta analyses in each data domain.
Then a decision is taken based on whether the ROPE is included in or excluded from a nominal uncertainty interval
(typically, with 95\% or 99\% probability).

\emph{Bayes factors}~\cite{Kruschke-bfactors,jeffreys-book,good-origin-bfactor} are another Bayesian technique that may be used, under certain conditions, to choose between models in a dichotomous fashion.
A Bayes factor is a \emph{ratio of likelihoods} $B_{1,2} = \like[d; h_1]/\like[d; h_2]$ of the same data $d$ under different hypotheses (that is, statistical models) $h_1$ and $h_2$.
In a nutshell, the Bayes factor $B_{1,2}$ expresses how much the data would shift the posterior probability in favor of $h_1$ over $h_2$:
$B_{1,2} < 1$ indicates that $h_1$'s probability decreases, whereas $B_{1,2} > 1$ indicates that it increases.
Therefore, \emph{if $h_1$'s and $h_2$'s prior probabilities are about the same}, $B_{1,2}$ can be used to decide whether the data supports more strongly $h_1$ or $h_2$.
When the prior probabilities are not similar in value, their exact ratio must be known in order to meaningfully interpret a Bayes factor.

We mentioned ROPE analysis and Bayes factor as examples of techniques that may be appropriate in certain specific contexts.
However, we stress again that there is no substitute for a careful, bespoke posterior analysis---performed with
domain knowledge---if we want to get to sound conclusions grounded in practical significance.

\myparagraph{Tools.}
A number of powerful probabilistic languages, geared towards analyzing Bayesian statistical models, have been developed in the last decade.
The best known are JAGS,\footnote{\url{http://mcmc-jags.sourceforge.net/}} BUGS,\footnote{\url{https://www.mrc-bsu.cam.ac.uk/software/bugs/}} Turing,\footnote{\url{https://github.com/TuringLang}} and Stan\footnote{\url{https://mc-stan.org/}};
Stan is probably the most mature and actively maintained at the time of writing.
All of these tools input a statistical model---such as the one in~\eqref{eq:basic-lr}---including information about priors,
as well as the analyzed data;
they output a posterior distribution computed numerically as we described in \autoref{sec:bayesian-inference-algo},
which can be visualized, analyzed, and used to build derived models.
This process is central to every Bayesian data analysis.

These tools are normally used as back ends through convenient interfaces in the most common statistical analysis toolsets---such as
R,\footnote{\url{https://www.r-project.org/}} Python's Scipy,\footnote{\url{https://www.scipy.org/}} and Julia.\footnote{\url{https://julialang.org/}}.
The interface functions take care of importing and exporting data with the back ends,
and provide functionality that makes building and using common models (such as regressive ones) straightforward.
Another interesting tool worth mentioning is JASP\footnote{\url{https://jasp-stats.org/}} which supports both frequentist and Bayesian analyses under a single, user-friendly interface. This way, it can provide an accessible way to design Bayesian analyses that replace the use of better-known frequentist techniques.

\section{Threats to Validity and Limitations}
\label{sec:limitations}

Like every empirical study, this paper's reanalyses (\autoref{sec:testing} and \autoref{sec:rosetta}) may incur threats to validity.
Rather than just discussing them as traditionally done~\cite{experiments-book}, we illustrate
how using Bayesian statistics can generally help to mitigate them (\autoref{subsec:threats}) and to avoid other common problems in empirical software engineering (\autoref{subsec:smells}).
This section concludes with a mention of the limitations of Bayesian statistics (\autoref{subsec:limitations}).

\subsection{Threats to Validity}
\label{subsec:threats}

Using Bayesian statistics is unlikely to affect \emph{construct} validity, which has to do with whether we measured what the study was supposed to measure. Our reanalyses reused the same data that was produced in the original studies by \autotest and by \rosetta;
therefore, the same construct validity threats---and threat mitigation strategies---apply to the reanalyses in \autoref{sec:bayesian-testing} and \autoref{sec:bayesian-rosetta}.

\emph{Conclusion} validity depends on the application of appropriate statistical tests.
As we discussed throughout the paper, frequentist statistics are often hard to apply correctly in a way that is also functional to answering a study's research questions; Bayesian statistics, which are more transparent and straightforward to interpret, can certainly help in this respect.
Thus, conclusion validity threats are generally lower in our reanalyses than in the original analyses.

\emph{Internal validity} is mainly concerned with whether causality is correctly evaluated.
This depends on several details of experimental design that are generally independent of whether frequentist or Bayesian statistics are used. However, one important aspect of internal validity pertains to the avoidance of \emph{bias}; this is where Bayesian statistics can help, thanks to their ability of weighting out many different competing models rather than restricting the analysis to two rigid hypotheses (null vs.\ alternative hypothesis).
This aspect is particularly relevant for \autoref{sec:rosetta}'s reanalysis of Rosetta Code data,
where Bayesian modeling replaced the inflexible null-hypothesis testing.

Since they can integrate previous, or otherwise independently obtained, information in the form or priors,
and are capable of modeling practically relevant information in a rich way, 
Bayesian statistics can help mitigate threats to \emph{external validity}, which concern the generalizability of findings.
In particular, prior sensitivity analysis (described in \autoref{sec:role-of-priors})
assessed the (in)dependence of the statistical conclusions on the choice of priors in both case studies (\autoref{sec:autotest:bayes-priors} and \autoref{sec:rosetta:bayes:sensitivity}), thus giving a clearer picture of the generalizability of their findings.

\subsection{Analytics Bad Smells}
\label{subsec:smells}

Menzies and Shepperd~\cite{SmellsAnalytics} recently proposed a list of 12 ``bad smells''---indicators of poor practices in analytics research.
Let's outline how using Bayesian statistics can help avoid most of these pitfalls.

\begin{enumerate}
\item \textbf{Not interesting.}
  While whether a study is interesting mainly depends on its design and topic, Bayesian statistics make it easier to model quantities
  that are of direct practical relevance.

\item \textbf{Not using related work.}
  \autoref{sec:relatedWork} indicates that Bayesian statistics is hardly ever used in empirical software engineering research,
  suggesting that there is ample room for progress in this direction.

\item \textbf{Using deprecated or suspect data.}
  This poor practice is independent of the statistical techniques that are used;
  in our reanalyses, the original studies' data is of good quality to the best of our knowledge.

\item \textbf{Inadequate reporting.}
  We make available all raw data, as well as the analysis scripts, we used in our reanalyses to help reproducibility and further analyses.
  Note that the raw data was already released by the original studies~\cite{autotest-bib,rosetta}.

\item \textbf{Underpowered studies.}
  The Bayesian focus on type $M$ and type $S$ errors (\autoref{sec:type12-sm-errors}), and its prior sensitivity analysis (\autoref{sec:role-of-priors}), help assess the limitations of a study that really matter
  to achieve practical significance.

\item \textbf{$p<0.05$ and all that!}
  It should be clear that one of the main recommendations of this paper is to dispense with $p$-values and statistical hypothesis testing.

\item \textbf{Assumptions (e.g., of normality) in statistical analysis.}
  Bayesian statistical modeling makes assumptions visible, and hence harder to apply without justification.
  Sensitivity analysis is a very effective way of assessing the impact of different assumptions.

\item \textbf{No data visualization.}
  Visualization is a valuable practice regardless of the kind of statistical techniques that are being used.
  Since Bayesian analysis provides full posterior distributions (as opposed to scalar summaries), visualization is
  a natural way to inspect details of an analysis's outcome---as shown, for example, in \autoref{fig:mvr-effe-bayes-gauss}, \autoref{fig:mvr-effe-poisson-reduced}, and \autoref{fig:posteriors-fsharp-vs-three}.

\item \textbf{Not exploring stability.}
  The practice of sensitivity analysis (\autoref{sec:role-of-priors}) is precisely a form of robustness assessment.
\end{enumerate}

The three remaining ``smells'' (\emph{not tuning}, \emph{not exploring simplicity}, and \emph{not justifying choice of learners})
mainly pertain to details of how statistical inference is carried out.
Since all of our models are simple (conceptually, and to analyze using standard tools), these smells are not really relevant, but they might be if we used significantly more complex models.
In these cases too, frequentist statistics---with its rigid procedures and unintuitive results---does not seem to be at an advantage.

\subsection{Limitations of Bayesian Statistics}
\label{subsec:limitations}

This paper discussed at length the limitations of frequentist statistics.
Let us now outline some limitations of Bayesian statistics.

Experimental design and data quality are primary determinants of the value of every data analysis's results.
This is true when applying Bayesian statistics as much as it is when applying frequentist statistics.
In particular, Bayesian analysis techniques are not a silver bullet, and cannot fix a botched study design or data that is irrelevant or spurious.

In contrast to frequentist techniques such as null hypothesis testing---which can be used largely as a black box---applying
Bayesian statistical techniques requires some \emph{modeling} effort.
Modeling means more than just following a ready-made recipe, and should be based on an understanding of the data domain.
This does not mean that Bayesian statistics cannot be used for exploratory studies; in fact, the transparency of Bayesian statistical models
helps understand the role and impact of every assumption in our analysis, which involves more effort but is ultimately beneficial to achieving clear results.

Like with every powerful analysis technique, it takes some practice to become familiar with Bayesian techniques and tools,
and to fully appreciate how they can be applied to the kinds of problems a researcher encounters most frequently.
To help this process, \autoref{sec:guidelines} discussed some basic guidelines that should be applicable to a variety of common analyses.
We plan to come up with a more comprehensive set of guidelines in future work.

Very practical limitations of Bayesian statistics in empirical software engineering
may derive from the mere fact that frequentist statistics are currently dominating the research practice (see \autoref{sec:relatedWork}).
This implies that it may be harder to publish analyses based on Bayesian statistics
simply because the research community expects frequentist statistics (e.g., ``statistical significance'')
and is unfamiliar with Bayesian techniques as best practices.
We are convinced these challenges will be only temporary; and hopefully the research community will be receiving favorably
the usage of Bayesian statistics.
Some of Bayesian data analysis's fundamental features can be key to its practical acceptance---in particular, its focus on practical significance, and the flexibility and understandability of its models and techniques.

\section{Discussion}
\label{sec:conclusions}

Frequentist statistics remain the standard choice for data analysis in empirical software engineering, as it offers frameworks that are easy to apply and widely accepted as good practices.
This situation is unfortunate, because frequentist techniques, as applied in practice, are plagued
by a number of issues that undermine their effectiveness in practice:
\begin{itemize}
\item statistical hypothesis testing over-simplifies complex analyses
  by phrasing them as binary choices between a straw-man null hypothesis
  and an alternative hypothesis;
\item $p$-values are, in general, unsound to assess statistical significance;
 \item multiple tests further challenge null-hypothesis testing, and there is no consensus
   on how to deal with them in a way that is safe and consistent;
\item effect sizes and confidence intervals are useful measures,
  but can be surprisingly tricky to interpret correctly in terms of practically significant
  quantities.
\end{itemize}

In this paper, we illustrated concrete instances of the above issues from an empirical software engineering perspective.
By using Bayesian statistics instead, we demonstrated gains in the rigor, clarity,
and generality of the empirical analyses of \autotest and \rosetta:
\begin{itemize}
\item Bayesian analysis stresses precise modeling of assumptions, uncertainty, and domain features,
  which help gain a deep understanding of the data; \item applying Bayes' theorem gives a full \emph{posterior distribution} of each estimated parameter,
  which quantifies the imprecision of the estimate and supports both visual inspections and numerical analyses;
\item the problem of multiple testing is immaterial, both because there is no ``hypothesis testing'' as such
  and because Bayesian analysis's quantitative posteriors support reasoning about the actual sign and magnitude of an effect
  rather than about possible compounding of errors that may or may not matter in practice;
\item \emph{priors} naturally incorporate assumptions and expectations (including from previous studies), and help understand to what extent analysis results
  depend on specific assumptions;
\item the numeric simulation-based tools implementing Bayesian analysis techniques
  offer a great flexibility and support \emph{predictions} in a variety of complex scenarios.
\end{itemize}

Based on these results, we recommend using Bayesian statistics
in empirical software engineering data analysis.
Besides supporting nuanced analyses in a variety of domains,
they have the potential to step up generality and predictive ability,
thus making progress towards the ultimate goal of
improving software engineering practice in terms of effectiveness and efficiency.
By using informed priors based on previous, related work they can further help link several empirical software engineering studies together,
and show what we currently can conclude based on them taken as a whole. This could help develop fruitful scientific practices
where each research contribution builds on the others, and they collectively make progress towards settling important problems. By simulating real-world scenarios that are relevant for practitioners such practices could also help better connect software engineering research to practice---thus fostering a greater impact of the former on the latter.

\subsubsection*{Acknowledgments}
We would like to thank Jonah Gabry (Columbia University, USA),
for discussions regarding some of this paper's content.
The first author also thanks for interesting discussions
the participants to the EAQSE (Empirical Answers to Questions in Software Engineering) workshop held in Villebrumier, France in November 2018,
where he presented a preliminary subset of this article's content.

\end{document}